\documentclass[%
 reprint,
superscriptaddress,
 amsmath,amssymb,
 aps,
prb,
]{revtex4-2}

\usepackage{graphicx}
\usepackage{dcolumn}
\usepackage{bm}

\usepackage{xcolor}
\usepackage[normalem]{ulem}
\begin{document}

\preprint{APS/123-QED}

\title{First-principles derived force field for h-BN monolayer nanostructures:\\
Applications to sheets, nanotubes and nanotori}

\author{Aristotelis P. Sgouros}
\affiliation{Theoretical and Physical Chemistry Institute, National Hellenic Research Foundation, Vass. Constantinou 48, 11635 Athens, Greece}
\email{asgouros@eie.gr}
\affiliation{School of Chemical Engineering, National Technical University of Athens (NTUA), Athens 15780, Greece}
\author{Markos Arapchatzis}
\affiliation{School of Physics, Department of Solid State Physics, Aristotle University of Thessaloniki, Thessaloniki 54124, Greece}%
\author{\\Nektarios N. Lathiotakis}
\affiliation{Theoretical and Physical Chemistry Institute, National Hellenic Research Foundation, Vass. Constantinou 48, 11635 Athens, Greece}%
\author{Konstantinos Papagelis}
\affiliation{School of Physics, Department of Solid State Physics, Aristotle University of Thessaloniki, Thessaloniki 54124, Greece}%
\affiliation{Institute of Chemical Engineering Sciences, Foundation for Research and Technology- Hellas (FORTH/ICE-HT), Patras 26504, Greece}%
\author{George Kalosakas}
\affiliation{Department of Materials Science, University of Patras, Patras 26504, Greece}

\date{\today}

\begin{abstract}

In this work, we present an empirical force field for hexagonal boron nitride (hBN) monolayers, derived via a bottom-up strategy from first principles calculations. We aim to deliver a simple analytical force field for boron nitride which is efficient and applicable to large-scale simulations, without compromising its accuracy. The force field is developed with the goal to analytically reproducing the potential energy contributions arising from planar bond-stretching and bond-angle-bending deformations, as well as from out-of-plane torsional deformations, as obtained from periodic density functional theory calculations. Analytical anharmonic potential energy functions were employed to describe and parameterize the force field through a fitting process. The potential is applied for estimating in-plane (stiffness matrix and elastic constants) and out-of-plane (bending and Gaussian rigidity) mechanical properties of monolayer hBNs using both molecular mechanics calculations and analytical formulas. Additionally, we determined the elastic properties of more complex nanostructures, including hBN nanotubes and nanotori. By combining a continuum model for bent hollow beams with the definition of bending rigidity, we validate the model predictions against an analytical expression for the bending rigidity of large-diameter hBN nanotubes, expressed as a function of their radius and the in-plane Young modulus.
\end{abstract}

\maketitle


\section{\label{sec:level1}Introduction}

Hexagonal boron nitride (hBN) has sparked significant interest within the scientific and industrial communities due to its unique properties. Being a two-dimensional material with a wide indirect bandgap (5.955 eV)\cite{Cassabois2016}, excellent chemical and thermal stability,\cite{JIANG2015589} exceptional thermal conductivity\cite{doi:10.1126/sciadv.aav0129} and good dielectric and optical properties,\cite{HUI2016119,10.1063/1.4764533}, it is highly versatile for numerous applications.\cite{C7TC04300G,JIANG2015589,10.1063/1.4764533,Molaei2021} In the field of electronics, hBN has been successfully employed in transistors, integrated circuits, and capacitors, leading to enhanced performance and improved energy conversion efficiency.\cite{Bao2016,Lee_2018,https://doi.org/10.1049/hve.2020.0076} Moreover, its remarkable thermal management capabilities have been utilized in electronic devices and green LEDs, effectively dissipating heat.\cite{choi2019application} The optical properties of hBN have also proven valuable in optoelectronics and photonics,\cite{caldwell2019photonics,deb2021boron} where it can serve as a transparent electrode, UV light emitter,\cite{LIU2021291} and as UV-detector.\cite{10.1063/1.4764533} In addition, hBN has been considered for applications in photocatalytic carbon dioxide reduction.\cite{D2TA09564E} Furthermore, hBN's chemical stability and lubricating properties have contributed to its use in protective coatings and high-temperature lubrication.\cite{saji20232d, TALIB2017360,chkhartishvili2016hexagonal} 

Besides hBN, various allotropic forms have been considered, including nanocages,\cite{Stephan1998} nanotubes,\cite{tamilkovan2024current,D2TA09564E} nanotori,\cite{10.1063/1.4827866} and nanoporous materials,\cite{JIANG2015589,weber2017boron} which have also been explored for diverse applications.
Ongoing research in this field continues to fuel scientific and technological interest in hBN, expanding its potential applications and broadening its impact across various disciplines.

The mechanical properties of hBN have been investigated in the literature via experiments,\cite{SOLOZHENKO19951, PhysRevB.73.041402} ab-initio calculations,\cite{Kudin2001, PENG201211,andrew2012mechanical,wu2013mechanics, Milowska2013} and atomistic molecular dynamics (MD) simulations using Tersoff-based force fields in most cases.\cite{MORTAZAVI20121846, Han_2014, Thomas_2016, Seremetis2020, PhysRevB.96.184108} In this work, we develop a semi-empirical force field for hBN monolayers encompassing both in-plane (bond-stretching and bond-angle-bending) and out-of-plane (torsional) interactions. The force-field parameterization results from a bottom-up fitting procedure in terms of matching the potential energy of properly deformed hBN geometrical structures as derived from first-principles density functional theory (DFT) calculations.

The functional expressions of the proposed potential are simple yet effective, capable of efficiently addressing very large hBN sheets with large-scale molecular dynamics simulations. The resulting force field is significantly more efficient than previous Tersoff-based implementations\cite{MORTAZAVI20121846, Han_2014, Thomas_2016, Seremetis2020, PhysRevB.96.184108}. It has been employed to investigate the structural relaxation and elastic properties of various hBN-based nanostructures, including hBN monolayers, nanotubes (NT), and hBN nanotori.

In the following, we provide a detailed description of the parametrization of the in-plane (Section \ref{in-plane ff}) and out-of-plane (Section \ref{out-of-plane ff}) terms of the force field through the fitting of the potential energy of deformed hBN monolayers obtained from ab-initio calculations. Subsequently, in Section \ref{In-plane Elastic} we determine the elastic constants of single-layer (SL) hBNs at \textit{T} = 0 K,  such as the components of stiffness matrix and the corresponding Lam\'e parameters, Young modulus, Poisson ratio, and bulk modulus. The aforementioned parameters are estimated via molecular mechanics calculations and with analytical closed-form formulas.\cite{Berinskii2010} In Section \ref{Out-of-plane Elastic}, we illustrate the variation of the potential energy of hBN nanotubes as a function of their radius and chirality. In the limit of infinite radius, we extrapolate the bending rigidity of the corresponding hBN monolayers to provide comparisons against analytical expressions. Finally, in Section \ref{PeTorus} following a similar procedure for the energy of folded hBN NT in order to form closed nanotori, the bending rigidity of hBN nanotubes is estimated.

\section{\label{Force Fields}Force Field Derivation}

We conducted a series of first principles', periodic DFT calculations for specifically deformed single-layer hBN structures to compute the energy variation upon changing a particular geometrical deformation parameter (Fig.~\ref{manipulations}).
The calculations were performed with the Quantum Espresso\cite{Giannozzi_2009,Giannozzi_2017,10.1063/5.0005082} code, employing the GGA/PBE functional\cite{PhysRevLett.77.3865} and the PAW method for the treatment of core electrons.\cite{PhysRevB.50.17953}
In all calculations, cut-offs for the wave functions and charge density were set at 60 Ry and 480 Ry, respectively.
Five different unit cells were considered; two for investigating in-plane deformations and three for the out-of-plane ones.

We considered the minimal 2-atom unit cell to simulate the in-plane deformations with an $8\times 8\times 1$ Monkhorst pack sampling of the reciprocal space. Benchmark calculations demonstrate that a \textit{k}-point mesh with $k \ge 8$ along the periodic directions yields converged forces for the systems under consideration. To determine the bond-stretching potential term, the unit cell lattice vectors were scaled uniformly (Fig.~\ref{manipulations}a). 
Regarding the bond-angle-bending term, the unit cell was deformed in such a way (Fig.~\ref{manipulations}b) that all bond lengths assumed rigid and only the angles were varied in a symmetric fashion (see Fig.~\ref{inset} below).

For the out-of-plane deformations, we considered deformed BN nanoribbons that were bent around a particular middle axis. Three different cases were considered:
(i) bending an 18 atom unit cell around a zig-zag (ZZ) middle axis with only boron atoms lying on that axis (Fig.~\ref{manipulations}c top, where the rotational zig-zag axis is shown by the green dotted line denoted as ZZ),
(ii) similarly bending a nanoribbon, with a 18 atom unit cell as well, around a zig-zag middle axis passing strictly through nitrogen atoms, and
(iii) bending a nanoribbon, with a 22 atom unit cell, around an armchair (AC) middle axis (Fig.~\ref{manipulations}c bottom, where the rotational armchair axis is shown by the green dotted line denoted as AC). For the out-of-plane deformations, at least an $1\times 24\times 1$ Monkhorst pack sampling of the reciprocal space was used. In all cases, a vacuum gap larger than 2 nm was maintained along the aperiodic directions to prevent spurious interactions.

\subsection{\label{in-plane ff}In-plane terms of the force field}

\begin{figure}[h]
\centering
  \includegraphics[width=8.25cm]{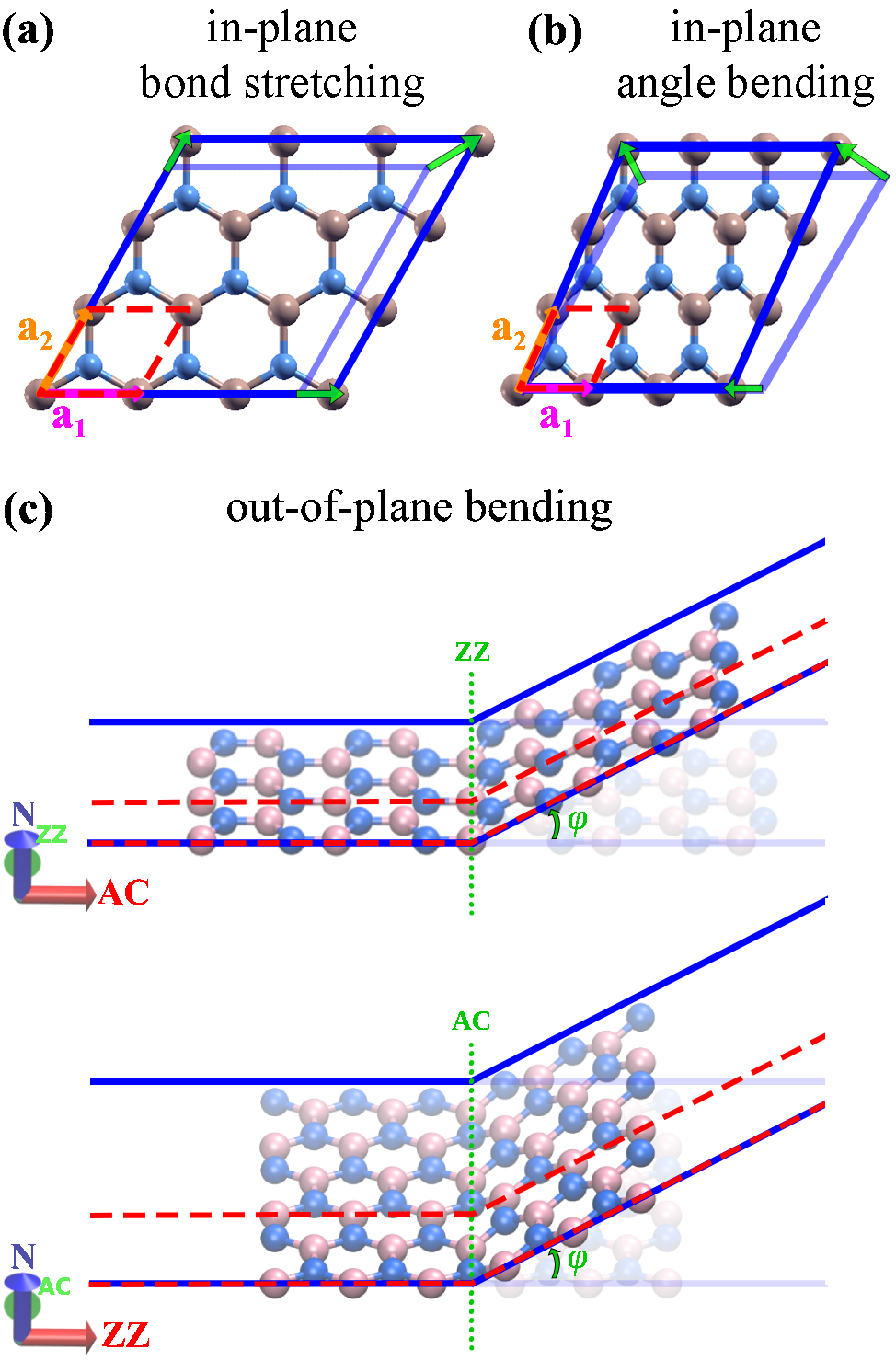}
  \caption{Schematic illustration of the deformed hBN structures considered in the DFT calculations. \textit{In-plane deformations}: \textbf{(a)} affine deformation with a uniform scaling along lattice vectors (Eq. (\ref{deformbondeq})) and \textbf{(b)} deformation altering in-plane angles and maintaining bond lengths (Eq. (\ref{deformangleeq})). Green arrows indicate the displacement direction of box corners.
  \textit{Out-of-plane deformations}: \textbf{(c)} folding an hBN ribbon by an angle $\varphi$ around a rotation axis (green dotted line) oriented along the ZZ (top) or AC (bottom) direction passing through its plane. In all cases, red dashed lines depict the edges of the unit cell used in the calculation. In (c), the box length in the aperiodic direction is much larger than the ribbon width to prevent spurious interactions; consequently, the unit cell edges normal to this direction are not shown. Supercells with three-unit cell replications along the periodic directions are drawn for clarity. The blue transparent (opaque) lines illustrate the periodic boundaries of the supercells before (after) the deformation.}
  \label{manipulations}
\end{figure}

The in-plane potentials of hBN were determined by deforming a periodic primitive cell with the following (unperturbed) lattice vectors: $\mathbf{a}_{1} = (a,0)$ and $\mathbf{a}_{2} = (\frac{a}{2},\frac{a\sqrt{3}}{2})$, where $a=\|\mathbf{a}_{1}\|=\|\mathbf{a}_{2}\|=\sqrt3 \; l_\textrm{BN}$ is the lattice constant and $l_\textrm{BN}$ the B-N bond length.
In the reference configuration ($\mathbf{R}^\textrm{ref}$), where there is no residual stress, the bond length equals 
 $l_\textrm{BN}=1.4518$ \AA~ and the lattice constant is $a=2.5146$ \AA; the error for $l_\mathrm{BN}$ and $a$ is lower than the reported significant digits.
 
The deformed configuration $\mathbf{R}^\textrm{def}$ was obtained by imposing a transformer $\mathcal{F}$ to the reference configuration
\begin{equation}\label{affineeq}
    \mathbf{R}^\textrm{def}=\mathcal{F} \cdot \mathbf{R}^\textrm{ref}\,.
\end{equation}
To extract the bond-stretching dependence of the deformation energy, we scaled the hBN lattice by applying the linear transformer 
\begin{equation}\label{deformbondeq}
    \mathcal{F}= \begin{bmatrix}
\epsilon & 0         \\
0        & \epsilon  \\ 
\end{bmatrix} + I_{2}\,,
\end{equation}
with $\epsilon$ being the engineering strain and $I_{2}$ the identity matrix of size 2; see illustration in Fig. \ref{manipulations}a. This operation scales the bond lengths uniformly and does not modify the in-plane angles. The same method has been applied in Ref.~\onlinecite{10.1063/1.4798384} for the case of graphene monolayers. 

\begin{figure}[h]
\centering
  \includegraphics[height=7.5cm,width=9.5cm]{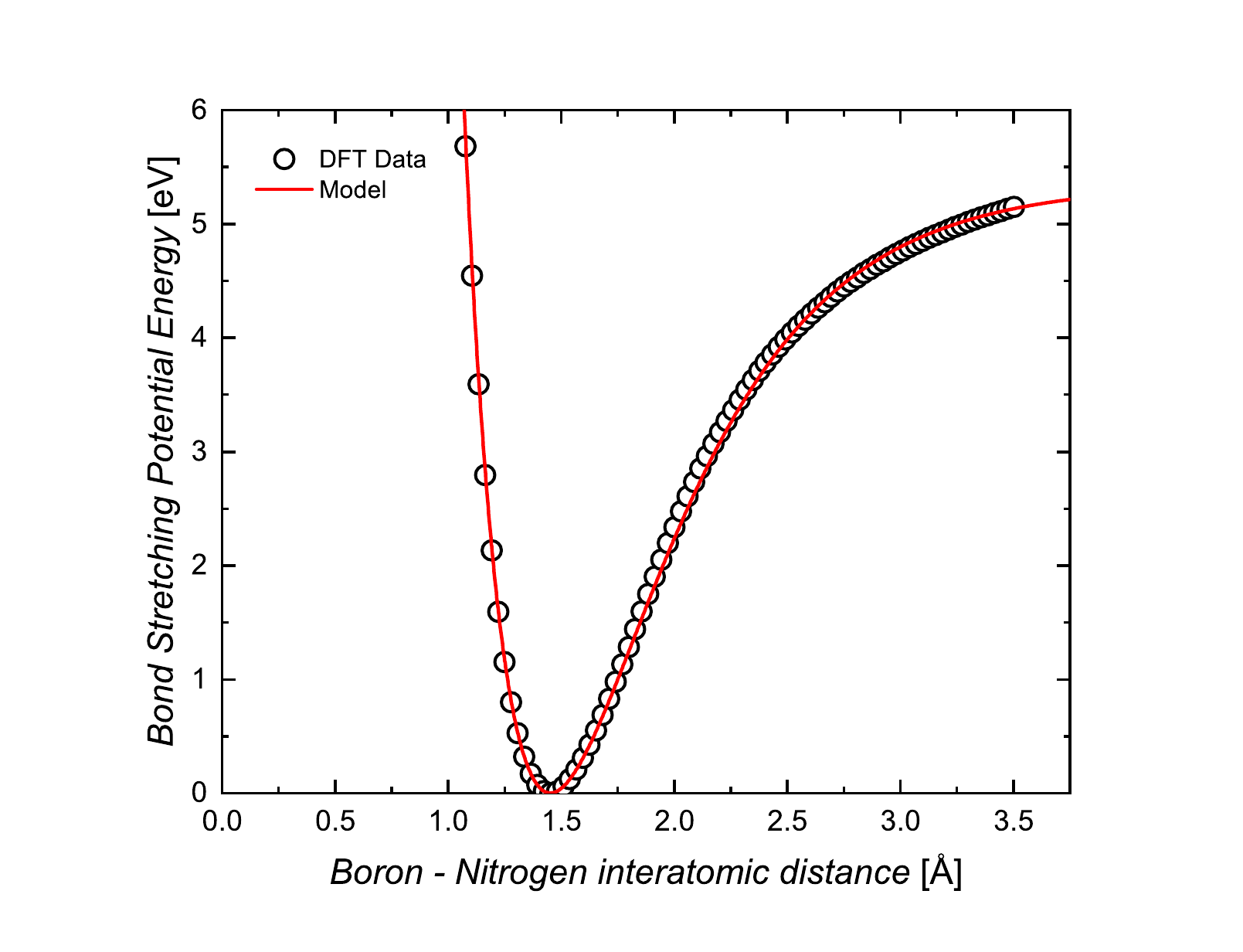}
  \caption{Bond-stretching energy as a function of the interatomic distance between a boron and a nitrogen atom. Black circles represent the DFT data. The solid line expresses the best fit with Eq.~(\ref{Morse}).}
  \label{bs}
\end{figure}

The bonded interactions of the force field were then parameterized through fitting of the DFT energy per bond results obtained from the uniform scaling deformations, shown in Fig.~\ref{bs} by circles, with the Morse potential
\begin{equation}\label{Morse}
    V_\textrm{b}(r)=D_\textrm{b}[e^{-\alpha_\textrm{b}(r-r_\textrm{b})}-1]^2\,,
\end{equation}
where $D_\textrm{b}$ is the potential well depth, $r_\textrm{b}$ the equilibrium bond distance (fixed parameter), and $\alpha_\textrm{b}$ the inverse width parameter of the potential. The fitting of the DFT data with Eq.~(\ref{Morse}) is shown by the continuous line in Fig.~\ref{bs}, yielding the following parameter values: $D_\textrm{b}=5.35 \pm 0.01$ eV, $\alpha_\textrm{b}=1.897 \pm 0.004$ \AA$^{-1}$, and $r_\textrm{b}=1.4518$ \AA. The calculated equilibrium B-N distance leads to a lattice constant $a=2.5146$ \AA~ 
in excellent agreement with literature values \cite{SOLOZHENKO19951,Paszkowicz2002,Thomas_2016,PENG201211,andrew2012mechanical,wu2013mechanics}.

\begin{figure}[h]
\centering
  \includegraphics[width=0.4\columnwidth]{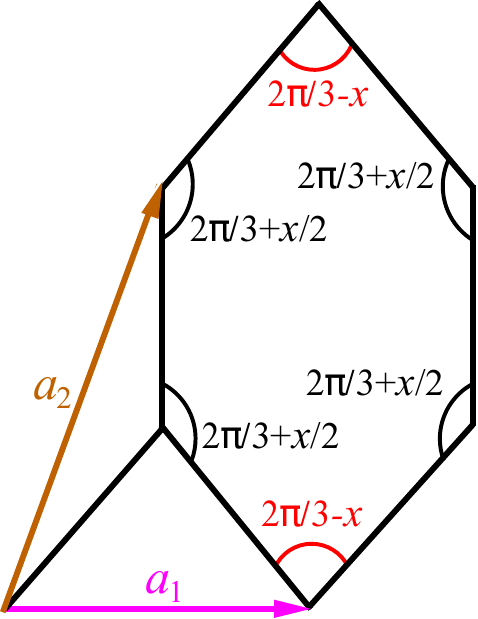}
  \caption{Schematic illustration of the angles in a deformed hBN ring according to the deformation shown in Fig.~\ref{manipulations}b.}
  \label{inset}
\end{figure}

The bond-angle-bending term of the force field is obtained through alteration of the in-plane angles of a pristine hBN, by $-x$ or $x/2$ while maintaining the equilibrium value of $1.4518$ \AA~in all B-N bonds, as shown in Fig.~\ref{inset}. \cite{10.1063/1.4798384}
In this case, four angles within each hBN ring increase by $x/2$, and two angles decrease by $x$. 
This operation deforms the reference configuration nonaffinely. First, we set the origin of the first (1) and second (2) sublattice (e.g., the pink and blue atoms in Fig.~\ref{manipulations}b, respectively) to

\begin{equation*}
 \begin{gathered}
        \mathbf{r}_{1}^\mathrm{ref}=\left(0,0\right) \,,\\
        \mathbf{r}_{2}^\mathrm{ref}=a\left(\frac{1}{2},\frac{\sqrt{3}}{2}\frac{C_x}{C_{x}+1}\right)\,,
\end{gathered}
\end{equation*}
and then, we apply the following linear transformer to the whole system
\begin{equation}\label{deformangleeq}
    \mathcal{F}= \begin{bmatrix}
\frac{2}{\sqrt{3}}\sqrt{1-{C_x}^2} & 0                                                        \\
0                                                   & \frac{2}{3}[1+C_x]         \\ 
\end{bmatrix}\,,
\end{equation}
where
\begin{equation}\label{Cxeq}
    C_x=\cos\left(\frac{\pi}{3}-\frac{x}{2}\right)\,.
\end{equation}

The energy contributions due to the deformation of the bond-bending angles were described by a nonlinear potential, containing quadratic and cubic terms
\begin{equation}\label{bendeq1}
    V_\textrm{a}(\phi)=\frac{k_\textrm{a}}{2}\Bigl(\phi-\phi_\mathrm{a}\Bigr)^2-\frac{k'_\textrm{a}}{3}\Bigl(\phi-\phi_\mathrm{a}\Bigr)^3\,,
\end{equation}
where $\phi_\mathrm{a} = \frac{2\pi}{3}$. Note that, Eq.~(\ref{bendeq1}) describes the energy of each in-plane angle in hBN.
The potential energy of a deformed hBN ring as that shown in Fig.~\ref{inset} equals to
\begin{equation}\label{ringeq}    V_\textrm{a}^\textrm{ring}(x)=2V_\textrm{a}\Bigl(\frac{2\pi}{3}-x\Bigr)+4V_\textrm{a}\Bigl(\frac{2\pi}{3}+\frac{x}{2}\Bigr)\,.
\end{equation}
Substituting Eq.~(\ref{bendeq1}) in Eq.~(\ref{ringeq}), allows to  express the potential energy of the deformed ring as a function of $k_\textrm{a}$ and $k'_\textrm{a}$: 
\begin{equation}\label{bendeq2}
    V_\textrm{a}^\textrm{ring}(x)=\frac{3}{2}k_\textrm{a}x^2+\frac{1}{2}k'_\textrm{a}x^3\,.
\end{equation}

\begin{figure}[h]
\centering
  \includegraphics[height=7.5cm,width=9.5cm]{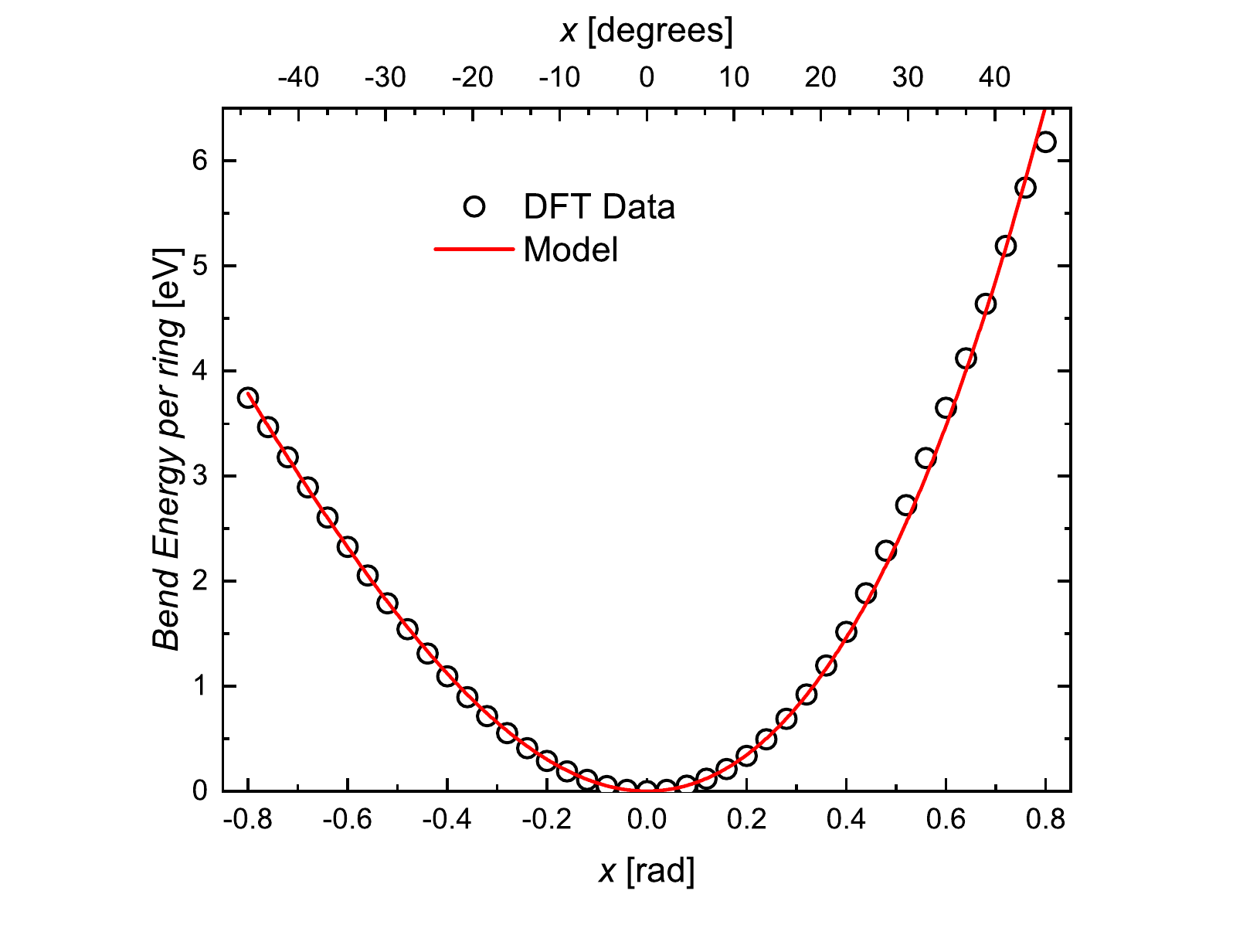}
  \caption{Total bond-angle-bending energy per ring as a function of the deformation parameter $x$, as shown in Fig.\ref{inset}. Hollow circles represent the DFT results. Solid line shows the fitting with Eq.~(\ref{bendeq2}).}
  \label{bend}
\end{figure}

Through fitting of the DFT energies per ring, corresponding to the deformations depicted in Fig.~\ref{manipulations}b, shown in Fig.~\ref{bend} by circles, with Eq.~(\ref{bendeq2}), the parameter values $k_\textrm{a}=5.37 \pm 0.02$ eV/rad$^2$ and $k'_\textrm{a}=5.35 \pm 0.01$ eV/rad$^3$ are obtained.

\subsection{\label{out-of-plane ff}Out-of-plane terms of the force field}

\begin{figure}[h]
    \centering
    \includegraphics[trim = 0mm 0mm 0mm 0mm, clip, width=0.4\linewidth]{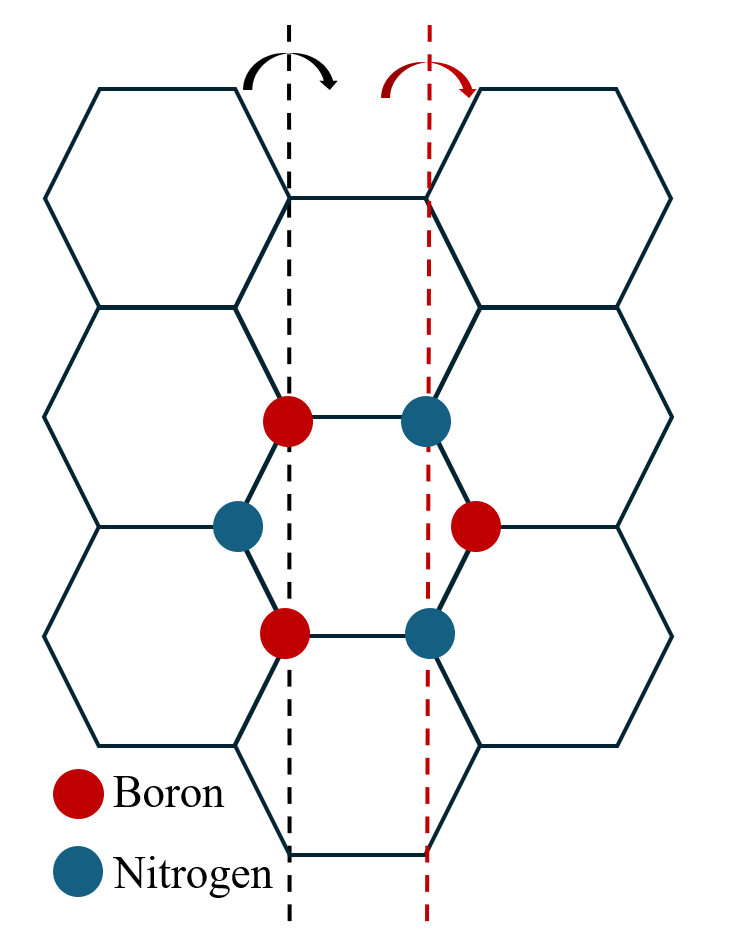}
    \includegraphics[trim = 0mm 0mm 0mm 0mm, clip, width=0.54\linewidth]{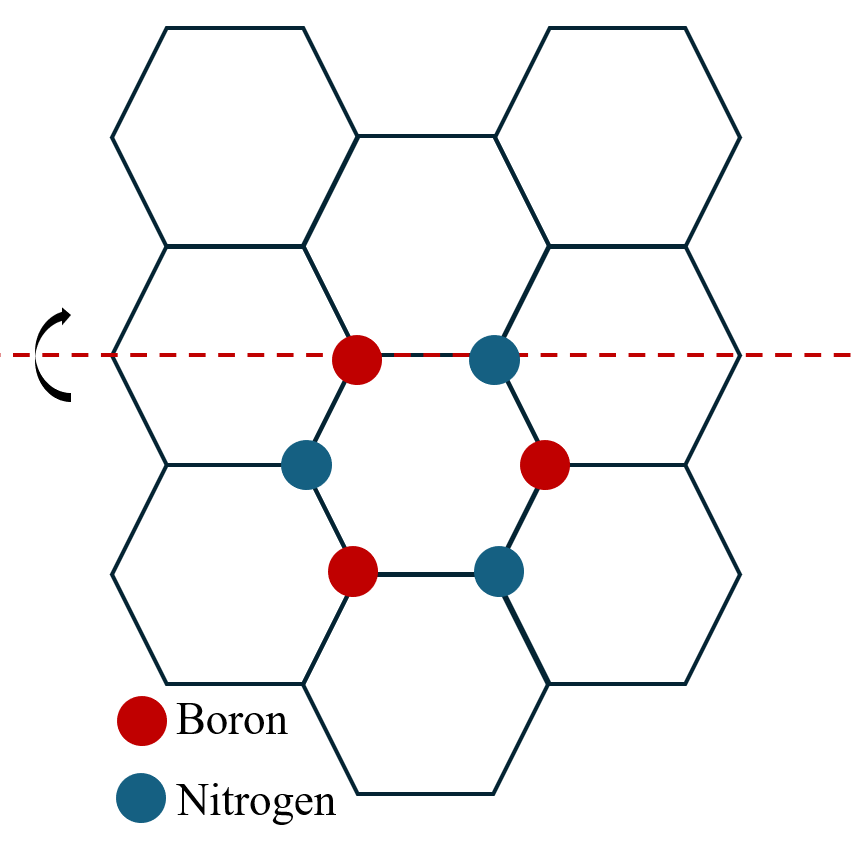}
    \caption{Schematic representation of a hBN honeycomb structure where the folding axes along the ZZ (left) or AC (right) directions are indicated. In the former case, two folding axes are shown, passing through second neighboring either B or N atoms, respectively. The structure is periodic along the folding axis in all cases.}
    \label{figex}
\end{figure}

To obtain the torsional terms of the force field, we determined through DFT calculations the variation of the potential energy of hBN nanoribbons by folding them along the zigzag or armchair directions by an angle $\mathit{\phi}$ (as illustrated in  Figs.~\ref{manipulations}c and~\ref{figex}). Under this deformation, the coordinates of the atoms on the right of the rotation axis in  Fig.~\ref{manipulations}c are transformed according to 
\begin{equation}\label{rotateeq}
    \begin{bmatrix} r^\textrm{def}_{\|} \\ r^\textrm{def}_{\perp} \\  
    \end{bmatrix} = 
    \begin{bmatrix}
        \cos(\phi) & -\sin(\phi)  \\
        \sin(\phi) & \cos(\phi)  \\
    \end{bmatrix} 
    \cdot
    \begin{bmatrix} r^\textrm{ref}_{\|} \\ r^\textrm{ref}_{\perp} \\
    \end{bmatrix} \,,
\end{equation}
with ${\|}$ denoting the direction parallel to the original ribbon plane and orthogonal to the rotation axis, and $\perp$ the direction normal to the ribbon plane. We assume that the origin of this coordinate system lies on the rotation axis and that the system retains its periodicity along the folding-axis direction. For example $(r_{\|},r_{\perp}) = (r_\textrm{AC},r_\textrm{N})$ in Fig. \ref{manipulations}c (top) and $(r_{\|},r_{\perp}) = (r_\textrm{ZZ},r_\textrm{N})$ in Fig. \ref{manipulations}c (bottom). Obviously, before the deformation, $r^\textrm{ref}_{\perp}=0$.
Concerning the finite ribbon widths (vertically to the folding axis, along the $r_{\|}$ direction), we considered values $\sim18.1$ \AA~ and $\sim12.6$ \AA~ for the ZZ and AC cases, respectively;
see the corresponding illustrations in Fig.~\ref{manipulations}c top and bottom. In order to extract the contribution of the torsional terms of the force field in these deformations and then to fit the relevant parameters of the out-of-plane potentials, we follow a procedure similar to that used in our previous work\cite{C7CP06362H, Chatzidakis2018} for single-layer graphene.

\begin{figure}[h]
\centering
  \includegraphics[height=7.5cm,width=9.5cm]{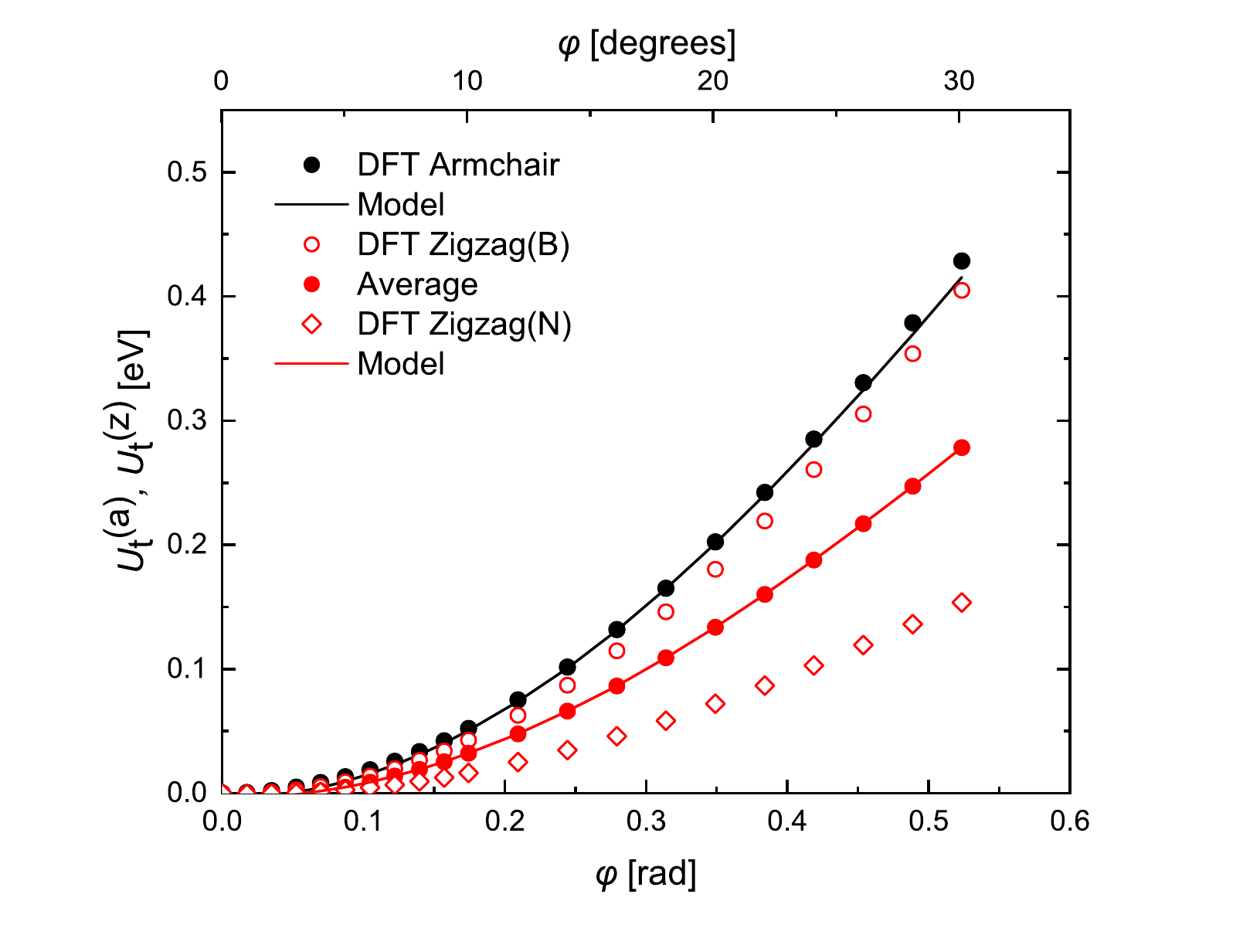}
  \caption{Torsional energy of hBN nanoribbons folded along AC (filled black circles) or ZZ (red open symbols) rotation axes. In the latter case, the results are shown by empty diamonds when the ZZ axis passes through N atoms and by empty circles when the axis passes through B atoms, as it is illustrated in Fig.~\ref{figex} (left). The average of these two situations (filled red circles) is considered in the fitting procedure regarding the rotation around ZZ folding axis.
  Solid lines indicate the model's fittings using Eqs. (\ref{torsionaleq1}) and (\ref{torsionaleq2}) for nanoribbons folded along the AC and ZZ axes.  }
  \label{Torsiong}
\end{figure}

Note that hBN exhibits mirror symmetry with respect to normal planes passing through neighboring B-N atoms; hence, a single deformation suffices for determining folding around the AC direction. The torsional energy of the AC-folded hBNs, $U\mathrm{_{t}^{(a)}}$ is shown in Fig.~\ref{Torsiong} by black circles.
On the contrary, hBN is asymmetric with respect to normal planes passing through second neighboring identical (B or N) atoms. To capture this effect in the case of folding around the ZZ direction, the torsional energy was averaged from two separate folding deformations, where the rotational axis passes through second neighboring either B or N atoms (see Fig.~\ref{figex} left).
The corresponding torsional energies $U\mathrm{_{t}^{(z)}}$ for the two separate ZZ folding deformations are shown by red open symbols in Fig.\ref{Torsiong}, while their average is represented by filled red circles.

Note that the total deformation energy obtained by DFT in these cases includes contributions not only from torsional terms but also from bond-angle-bending terms. Before fitting the parameters of the out-of-plane potentials one must remove the latter contributions, in order to isolate the dependence of the pure torsional terms on the folding angle $\mathit{\phi}$. 
Details of this fitting procedure can be found in Ref.~\onlinecite{Chatzidakis2018}.

As it is customary, we define the torsional angle, $\omega$, for a quadruple of atoms $(ijkl)$ as the dihedral angle of the planes defined by $(ijk)$ and $(jkl)$ atom triplets. One can distinguish two kinds of such angles, termed cis and trans. As in ref.~\onlinecite{Chatzidakis2018}, we used the following analytic formulas for the out-of-plane distortions as functions of  $\mathit{\omega}$:
\begin{equation}\label{torsionaleq1}
    V_\mathrm{t}^\mathrm{cis}(\omega)=k_\mathrm{t}^\mathrm{cis}[1-\cos(2\omega)] \,,\\
\end{equation}
\begin{equation}\label{torsionaleq2}
    V_\mathrm{t}^\mathrm{trans}(\omega)=k_\mathrm{t}^\mathrm{trans}[1-\cos(2\omega)].
\end{equation}
Fitting the torsional energy contribution obtained for the considered folding deformations using the analytic formulas of Eqs.~(\ref{torsionaleq1}) and (\ref{torsionaleq2}),  to the pure torsional energy extracted by the DFT data (when the bond angle bending terms have been substracted), the optimal parameter values $k_\mathrm{t}^\mathrm{cis}=0.066$ eV and $k_\mathrm{t}^\mathrm{trans}=0.109$ eV are derived. We have considered and fitted folding deformation data up to a rotational angle around $\mathit{\phi}=30$\textdegree.

\begin{table*}
  \caption{\ Parameters of the proposed force field for nanostructures of hBN monolayers.}
  \label{modeltable}
  \begin{center}
  \begin{tabular*}{0.95\textwidth}
  {@{\extracolsep{\fill}}lll}
    \hline\hline
    Interaction              & Potential Energy term   & Parameter values   \\[2pt]
    \hline
    Bond-stretching          & $V_\textrm{b}(r)=D_\textrm{b}[e^{-\alpha_\textrm{b}(r-r_\textrm{b})}-1]^2$  &  $D_\textrm{b}=5.35\textrm{ eV}$, $\alpha_\textrm{b}=1.897$ \AA$^{-1}$, $r_\textrm{b}=1.4518$ \AA  \\ [2pt]
    Bond-angle-bending       &    $V_\textrm{a}(\phi)=\frac{k_\textrm{a}}{2}\Bigl(\phi-\phi_\mathrm{a}\Bigr)^2-\frac{k'_\textrm{a}}{3}\Bigl(\phi-\phi_\mathrm{a}\Bigr)^3$ &   $k_\textrm{a}=5.37$ eV/rad$^2$, $k'_\textrm{a}=5.35$ eV/rad$^3$, $\phi_\mathrm{a} = \frac{2\pi}{3}$        \\[2pt]
    Torsional angles         &   $V_\mathrm{t}^\mathrm{cis/trans}(\omega)=k_\mathrm{t}^\mathrm{cis/trans}[1-\cos(2\omega)]$  &   $k_\mathrm{t}^\mathrm{cis}=0.066 \textrm{ eV}$, $k_\mathrm{t}^\mathrm{trans}=0.109 \textrm{ eV}$.        \\
    \hline\hline
  \end{tabular*}
  \end{center}
\end{table*}

To summarize the main results of this section, our proposed force field is given by the Eqs.~(\ref{Morse}) and (\ref{bendeq1}) for the in-plane bond-stretching and bond-angle-bending terms, respectively, and the Eqs.~(\ref{torsionaleq1}) and (\ref{torsionaleq2}) for the corresponding out-of-plane cis and trans torsional terms. The fitted parameters are shown in Table~\ref{modeltable}. 

Unlike graphene, hBN exhibits significant atomic charges, which are sensitive to structural deformations. As shown in Fig.~\ref{bend}, the DFT-calculated angle-bending energy exhibits a significant asymmetry around the equilibrium angle. This asymmetry arises from the increasing proximity of like-charged atoms as the bending angle, $x$ in Fig.~\ref{inset}, increases. The incorporation of these charges through additional fitted Coulomb terms is rather complicated. Even though explicit Coulomb terms are not included in our model, these interactions are implicitly captured within the fitted potentials. For instance, the aforementioned asymmetry in the angle-bending potential is described by the pronounced value of the parameter $k'_\mathrm a$ in the respecting nonlinear term. Similarly the contribution of the Coulomb interaction in the bond stretching term is implicitly considered through the fitting of the DFT-calculated energy.

The presented potentials can be efficiently used to simulate mechanical, dynamical, or other properties of hBN, as in the case of similar calculations in graphene \cite{Los2005,Fasolino2007,Neek-Amal2010,Xu2010,PhysRevB.86.125418,Sgouros2016,PhysRevB.95.035423,DAVINI201796,Sgouros2018,RAJ2019124,Genoese2019,SAVIN2020113937,Sgouros2021,KUMAR2021413250,ma15010067,PhysRevB.105.205414,PACHECOSANJUAN2023118416,Sgouros2024,2024IJMS..27309208V,PhysRevB.110.075434}.
In the following section, our force field is applied to determine elastic constants of hBN monolayers and nanotubes, using molecular mechanics calculations and analytical calculations.

\section{\label{Elastic Properties}Elastic Properties of hBN Nanostructures} 

\subsection{\label{In-plane Elastic}In-plane Elastic Constants of hBN Monolayers}

\begin{figure}[h]
\centering
  \includegraphics[height=6.87cm,width=8.25cm]{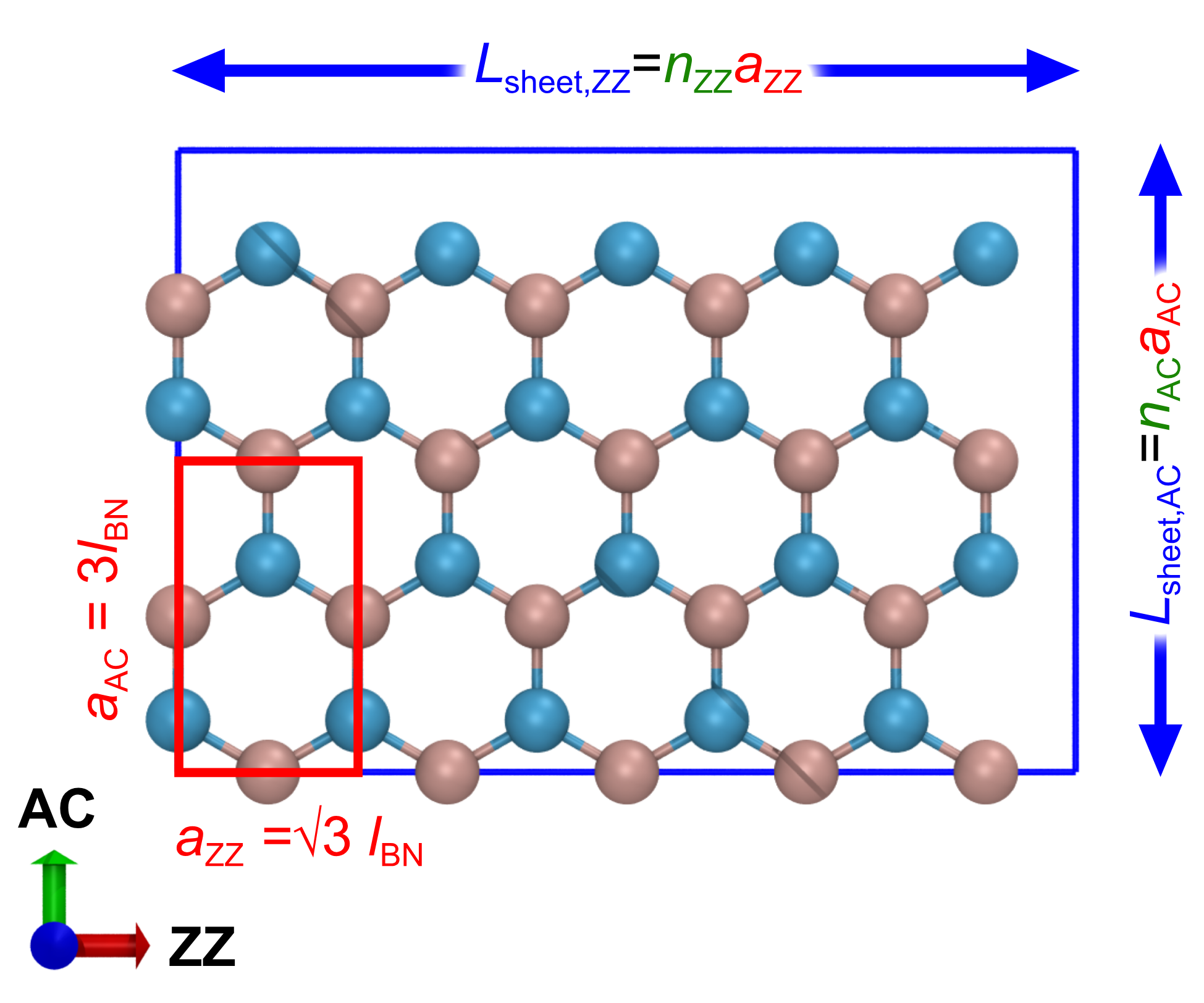}
  \caption{Schematic illustration of a periodic hBN SL constructed by replicating the 4-atom orthogonal cell (red rectangle) $n_\textrm{ZZ}=5$ and $n_\textrm{AC}=2$ times along the ZZ and AC directions, respectively. The cell has dimensions $a_\textrm{ZZ} = \sqrt{3} \; l_\textrm{BN}$ and $a_\textrm{AC} = 3 \;l_\textrm{BN}$. The vectors of the  orthorhombic simulation box are $\mathbf{A} = (n_\mathrm{ZZ}a_\mathrm{ZZ}, 0)$ and $\mathbf{B} = (0, n_\mathrm{AC}a_\mathrm{AC})$.  The blue rectangle depicts the periodic boundaries.}
  \label{schema_sheet}
\end{figure}

In this subsection, we will determine the elastic constants of periodic hBN single layers within the plane of the two-dimensional material, via molecular mechanics (MM) calculations by utilizing the force field developed in this work. The MM calculations were performed with the open source large-scale atomic/molecular massively parallel simulator (LAMMPS).\cite{Thompson2022}
The hBN sheets were generated upon replicating the 4-atom unit cell (see red rectangle in Fig. \ref{schema_sheet}) with dimensions $a_\textrm{ZZ} = \sqrt{3} \; l_\textrm{BN}$ and $a_\textrm{AC} = 3 \;l_\textrm{BN}$ along the ZZ and AC directions, $n_\textrm{ZZ}$ and $n_\textrm{AC}$ times, respectively.

The LAMMPS data files containing atomic coordinates and topology information (bonds, bond-bending angles, and torsion angles) were generated with the CrystalBuilder tool;\cite{Sgouros2024crystalbuilder} the latter has been extended in this work, in order to account for the different parameterization of the cis from the trans torsional angles.

The stiffness matrix $\mathit{\mathbf{C}}$ relates the components of strain tensor $\bm{\epsilon}$ and stress tensor $\bm{\sigma}$ according to 
\begin{equation}\label{ceq1}
    \bm{\sigma} = \bm{C}\bm{\epsilon} \\
\end{equation}
\noindent or 
\begin{equation}\label{ceq2}
    \begin{bmatrix} \sigma_{xx} \\ \sigma_{yy} \\ \sigma_{xy} \\ 
    \end{bmatrix} = 
    \begin{bmatrix}
C_{xxxx} & C_{xxyy} & C_{xxxy} \\
C_{yyxx} & C_{yyyy} & C_{yyxy} \\
C_{xyxx} & C_{xyyy} & C_{xyxy} \\
\end{bmatrix} 
    \begin{bmatrix} \epsilon_{xx} \\ \epsilon_{yy} \\ 2\epsilon_{xy} \\ 
    \end{bmatrix}\,,
\end{equation}
where pair indices involving the \textit{z}-components have been omitted restricting ourselves in the two-dimensional plane. Conveniently the stiffness matrix can be expressed in Voigt notation\cite{voigt1928lehrbuch} as 
\begin{equation}\label{ceq2a}
    \begin{bmatrix} \sigma_{1} \\ \sigma_{2} \\ \sigma_{6} \\ 
    \end{bmatrix} = 
    \begin{bmatrix}
C_{11} & C_{12} & C_{16} \\
C_{21} & C_{22} & C_{26} \\
C_{61} & C_{62} & C_{66} \\
\end{bmatrix} 
    \begin{bmatrix} \epsilon_{1} \\ \epsilon_{2} \\ \epsilon_{6} \\ 
    \end{bmatrix}\,,
\end{equation}
where now pair indices are denoted by single numbers, $xx=1$, $yy=2$, and $xy=6$. In doing so, the strain is described by the vector
\begin{equation}\label{strainvector}
\bm{\epsilon} = (\epsilon_{1}, \epsilon_{2}, \epsilon_{6}) = (\epsilon_{xx}, \epsilon_{yy}, 2\epsilon_{xy})\,.
\end{equation}

To determine the elements of the stiffness matrix via MM computations we utilize three fundamental deformation modes (elementary deformations), each characterized by a single non-zero strain component in Eq. \ref{strainvector}; hence, the components of the stiffness matrix are derived from the slope of the corresponding stress-strain plots in the linear regime for small strains.\cite{Clavier2017}

The first $\bm{\epsilon} = (\epsilon_{1}, 0, 0)$ and second $\bm{\epsilon} = (0,\epsilon_{2}, 0)$ deformation modes entail deforming the box by an elementary amount $\xi$ along the $x$ and $y$ directions by applying the linear transformers
\begin{equation*}
    \mathcal{F} = \begin{bmatrix}
        \xi & 0 \\
        0   & 0 \\
\end{bmatrix} + I_{2}
\qquad \mathrm{and} \qquad 
    \mathcal{F} = \begin{bmatrix}
        0   & 0   \\
        0   & \xi \\
\end{bmatrix} + I_{2}\,,
\end{equation*}
respectively; note that $\epsilon_{1} = \xi$ and $\epsilon_{2} = \xi$.
The third deformation mode $\bm{\epsilon} = (0, 0, \epsilon_{6})$ entails varying the angle $\hat{\gamma}$ between the lattice vectors of an initially orthorhombic box (shearing) upon applying the linear transformer
\begin{equation*}
    \mathcal{F} = \begin{bmatrix}
        0   & \xi \\
        0   & 0 \\
\end{bmatrix} + I_{2}\,,
\end{equation*}
where $\epsilon_{6} = \tan(\hat{\gamma}\xi)\approx \hat{\gamma}\xi$. The slope of the corresponding stress-strain plots was evaluated with a central-difference scheme with perturbation $\xi = \pm10^{-7}$. For additional details regarding the determination of the stiffness matrix with explicit deformation, the interested reader is referred to the article by Clavier et al;\cite{Clavier2017} see Section 2.1 therein. Details regarding the calculation of the stress tensor can be found in Ref. \onlinecite{10.1063/1.3245303}.

Henceforth, we will work with two-dimensional (2D) stress values which are related to the 3D stress as 
\begin{equation}\label{stress2d3deq}
    \mathcal{X} =  l_{0}\mathcal{X}^\mathrm{3d}\,,
\end{equation}
with $l_\mathrm{t}$ being the characteristic thickness of the monolayer, $\mathcal{X}$ a quantity with 2D stress units (Pa$\cdot$m) and $\mathcal{X}^\mathrm{3d}$ a quantity with 3D stress units (Pa). Care should be exercised when converting between 3D and 2D stress values, since various conventions have been invoked across the literature concerning the SL thickness.\cite{Thomas_2016,Han_2014,PhysRevB.73.041402,Milowska2013} The results of our calculations regarding the components of the stiffness matrix in units of 2D stress are summarized in Table \ref{elastictable}.

\begin{table*}
  \caption{\ The lattice parameter \textit{a}, the in-plane and out-of-plane (bending rigidity $D_\mathrm{sheet}$ and Gaussian rigidity $D_\mathrm{G,sheet}$) elastic constants of hBN monolayers, calculated through MM simulations using the proposed force field, along with comparisons of corresponding experimental and theoretical (atomistic or ab initio methods) values reported in the literature\textsuperscript{1}.
 }

  \label{elastictable}
  \begin{center}
  \begin{tabular*}{1.0\textwidth}{@{\extracolsep{\fill}}lllll}
    \hline\hline
    Parameter               & This work  & Experiment                          & Atomistic simulations                      & Ab initio \\
    \hline
    $a$ (\AA)               & 2.5146     & 2.504 \cite{SOLOZHENKO19951}, 2.505 \cite{Paszkowicz2002} 
                                                                               & 2.505 \cite{Thomas_2016,GovindRajan2018}, 2.504 \cite{PhysRevB.96.184108} 
                                                                                                                            & 2.512 \cite{PENG201211}, 2.51 \cite{andrew2012mechanical}, 2.504 \cite{wu2013mechanics}  \\
    $C_{11}, C_{22}$ (Pa$\cdot$m) & 279.3    & 270 \cite{PhysRevB.73.041402}       & 274.98 \cite{Thomas_2016}, 230.44 \cite{GovindRajan2018}, 277 \cite{PhysRevB.96.184108} 
                                                                                                                            & 293.2 \cite{PENG201211}, 289.8 \cite{andrew2012mechanical}  \\
    $C_{12} = \lambda$ (Pa$\cdot$m)         & 76.9     & 56.3 \cite{PhysRevB.73.041402}      & 81.91 \cite{Thomas_2016}, 50.28 \cite{GovindRajan2018} 
                                                                                                                            & 66.1 \cite{PENG201211},  63.7 \cite{andrew2012mechanical}  \\
    $C_{66} = \mu = G$ (Pa$\cdot$m)         & 101.2    & -                                   & 96.68 \cite{Thomas_2016}                   & 107.92 \cite{Milowska2013}, 113.1 \cite{andrew2012mechanical} \\
    $\nu$                   & 0.275      & 0.208 \cite{Boldrin_2011,PhysRevB.73.041402}, 0.200 \cite{Boldrin_2011,JagerPhD1977}                                   
                                                                               & 0.297 \cite{Thomas_2016}, 0.17 \cite{nano11113113, Mayo1990}
                                                                                                                            & 0.216 \cite{Milowska2013}, 0.2176 \cite{PENG201211} \\
                            &            &                                     & 0.16 \cite{nano11113113, Boldrin_2011}, 0.186 \cite{PhysRevB.96.184108}
                                                                                                                            &  \\
    $Y$ (Pa$\cdot$m)              & 258.2    & 288.9 $\pm$ 23.4\cite{Falin2017},   & 250.5 \cite{Thomas_2016}, 293.4 \cite{Han_2014}, 274.7 \cite{MORTAZAVI20121846},
                                                                                                                            & 262.3 \cite{Milowska2013}, 278.3 \cite{PENG201211}, 271\cite{Kudin2001} \\
                            &            & 256 \cite{Boldrin_2011,PhysRevB.73.041402}, 237 \cite{Boldrin_2011,JagerPhD1977}     
                                                                               & 295 \cite{nano11113113, Mayo1990}, 280 \cite{nano11113113, Boldrin_2011}, 267 \cite{PhysRevB.96.184108}
                                                                                                                            & 275.8 \cite{andrew2012mechanical}, 256.245 \cite{wu2013mechanics} \\
    $B$ (Pa$\cdot$m)              & 178.1    & -                                   & -                                          & 177 \cite{andrew2012mechanical} \\
    $D_\textrm{sheet}$ (eV) & 1.56      & -                                   & 0.56 \cite{Thomas_2016}, 0.86 \cite{PhysRevB.87.184106}, 0.89 \cite{YI2018408}
                                                                                                                            & 1.29 \cite{Kudin2001}, 0.95 \cite{wu2013mechanics} \\
    $D_\textrm{G,sheet}$ (eV) & $-3.23$ & -                                   & -                                          & - \\
    \hline\hline
  \end{tabular*}
    \end{center}
 \noindent{\footnotesize {\textsuperscript{1} The characteristic thickness ($l_\mathrm{t}$) for converting 3D stress to 2D units, Eq. (\ref{stress2d3deq}), has been considered 3.34 \AA~ 
 in Ref. \cite{Thomas_2016, Falin2017}, 3.3 \AA~ in Ref. \cite{wu2013mechanics}, 3.33 \AA~ in Refs. \cite{Han_2014,PhysRevB.73.041402,GovindRajan2018}, and $r_\mathrm{N}+r_\mathrm{B}=1.55$~\AA~$+1.92$~\AA~$=3.47$~\AA~ in Ref. \cite{Milowska2013} ($r_\mathrm{N}$ and $r_\mathrm{B}$ are the van der Waals radii of N and B atoms, respectively). }}

\end{table*}

The components of the stiffness matrix and the corresponding elastic constants are in good agreement with the values reported in the literature from MD and ab initio simulations. In addition, the lattice constant conforms with the reported experimental and theoretical estimates.

For isotropic 2D membranes, the stiffness matrix can be expressed in terms of the Lam\'e parameters $\lambda $ and $\mu$
\begin{equation}\label{stiffnesslameeq}
 \bm{C} =
    \begin{bmatrix}
\lambda+2\mu & \lambda      & 0 \\
\lambda      & \lambda+2\mu & 0 \\
0            & 0            & \mu \\
\end{bmatrix} \nonumber
\end{equation}
and the corresponding elastic constants of two-dimensional materials are obtained by
\begin{equation}\label{sheermoduluseq}
G=\mu\,,
\end{equation}
\begin{equation}\label{poissoneq}
\nu=\frac{\lambda}{\lambda+2\mu}\,,
\end{equation}
\begin{equation}\label{youngeq}
Y=\frac{4\mu(\lambda+\mu)}{\lambda+2\mu}\,,
\end{equation}
\begin{equation}\label{bulkmoduluseq}
B=\mu+\lambda\,,
\end{equation}
where $G$ represents the shear modulus, $\nu$ the Poisson ratio, $Y$ the Young modulus and $B$ the bulk modulus.

For validation purposes, the Lam\'e parameters have been evaluated analytically through the second derivatives of the in-plane potentials. In particular, denoting by $K''_\textrm{b}$ the second derivative of the bond-stretching potential, Eq.~(\ref{Morse}), at the equilibrium bond length
\begin{equation}\label{kappabondderveq}
    K''_\textrm{b} = \frac{d^{2}V_\mathrm{b}}{dr^{2}}\Bigr|_{r=r_\mathrm{b}} = 2\alpha_\mathrm{b}^2D_\mathrm{b}
\end{equation}
and $K''_\textrm{a}$ the second derivative of the bond-angle-bending potential, Eq.~(\ref{bendeq1}), at the equilibrium angle divided by the squared bond length
\begin{equation}\label{kappaanglederveq}
    K''_\textrm{a} = \frac{1}{{l_\mathrm{BN}}^2}\frac{d^{2}V_\textrm{a}}{d\phi^{2}}\Bigr|_{\phi=\phi_\textrm{a} } = \frac{k_\mathrm{a}}{{l_\mathrm{BN}}^2}\,.
\end{equation}
Then the Lam\'e parameters are provided by
\begin{equation}\label{lambdaanalytseq}
    \lambda=\frac{K''_\textrm{b}}{2\sqrt{3}}\frac{\tilde{K}-1}{\tilde{K}+1}\,,
\end{equation}
\begin{equation}\label{muanalytseq}
    \mu=\frac{K''_\textrm{b}}{\sqrt{3}}\frac{1}{{\tilde{K}+1}}\,,
\end{equation}
where
\begin{equation}\label{kappatildeeq}
    \tilde{K}=\frac{K''_\textrm{b}}{6K''_\textrm{a}}\,
\end{equation}
is a dimensionless parameter describing the relative stiffness of the bond-stretching and bond-angle-bending terms.

By applying Eqs. (\ref{lambdaanalytseq}) and (\ref{muanalytseq}) in Eq. (\ref{poissoneq}) we derive the following expression for the Poisson's ratio,
\begin{equation}
    \nu=\frac{\tilde{K}-1}{\tilde{K}+3}.
\end{equation}
Poisson's ratio depends strictly on $\tilde{K}$ and not on the individual stiffness of the bonds and bond-bending angles. Interestingly, when $\tilde{K} < 1$ Poisson's ratio becomes negative, resulting in an auxetic response.

Evaluating the above analytical expressions we see that the numerically derived Lam\'e parameters shown in Table \ref{elastictable} were obtained with an accuracy of the order of $(v_\mathrm{num}-v_\mathrm{analyt})/v_\mathrm{analyt} \sim 3\times10^{-7}$ with respect to the exact results, commensurate to the perturbation ($\xi = 10^{-7}$) applied in the finite difference scheme. It is worth mentioning that Eqs. (\ref{kappabondderveq})-(\ref{kappatildeeq}) are equivalent to the expressions derived
within the framework of linear elastic deformation by Berinskii and Krivtsov \cite{Berinskii2010} (compare for example with Eq.~5.4 therein).

\subsection{\label{Out-of-plane Elastic}Elastic Energy of single layer hBN Nanotubes and the Bending Rigidity of hBN Monolayers}

\begin{figure}[h]
\centering
    \includegraphics[width=7.5cm]{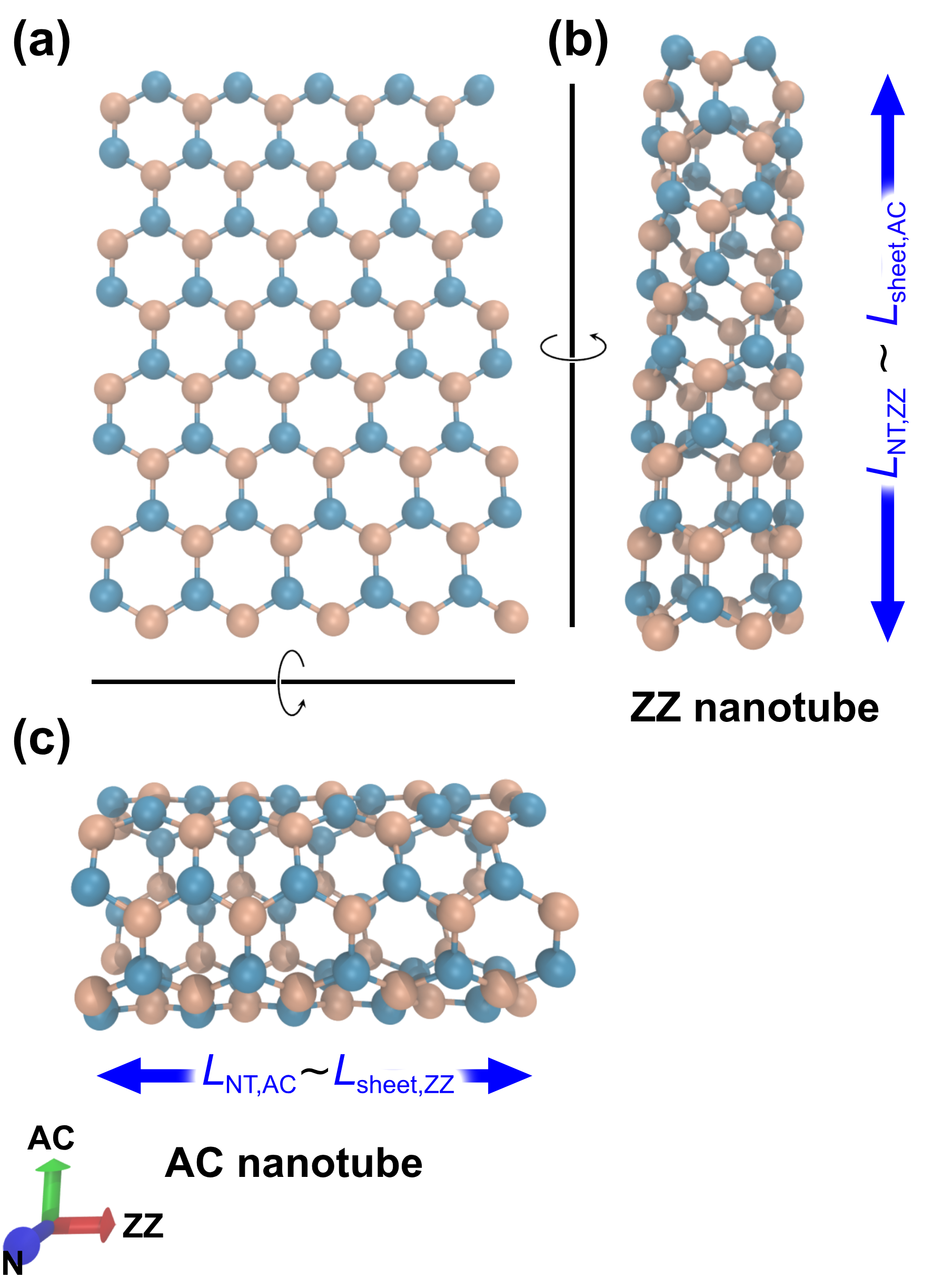}
  \caption{\textbf{(a)} Schematic illustration of an hBN monolayer with $(\textit{n}_\textrm{ZZ}, \textit{n}_\textrm{AC})$ = (5, 4). \textbf{(b)} Rolling the SL along the ZZ direction (vertical rotation axis) results in an ZZ nanotube with approximate length, $\textit{L}_\textrm{NT,ZZ} \sim \textit{n}_\textrm{AC}\textit{a}_\textrm{AC}$. \textbf{(c)} Similarly, rolling the SL along the AC direction (horizontal rotation axis) results in a AC nanotube with approximate length, $\textit{L}_\textrm{NT,AC} \sim \textit{n}_\textrm{ZZ}\textit{a}_\textrm{ZZ}$. 
  }
  \label{schema_nt}
\end{figure}
Here we present results for the elastic energy of AC and ZZ hBN nanotubes with varying radii. The NTs were generated by transforming the coordinates $x_{i}$, $y_{i}$, $z_{i}$ of each atom ($i$) of a periodic planar sheet with dimensions $L_{\mathrm{sheet},x} = n_{x}a_{x}$ and $L_{\mathrm{sheet},y} = n_{y}a_{y}$ according to
\begin{equation}\label{ntdeformeq}
 \begin{gathered}
 x_{\textrm{NT},i} = x_{i} \,,\\
 y_{\textrm{NT},i} = (R_{\mathrm{NT}}+\Delta z_{i})\sin\left(y_{i}/R_{\mathrm{NT}}\right) \,,\\
 z_{\textrm{NT},i} = (R_{\mathrm{NT}}+\Delta
 z_{i}) \cos\left(y_{i}/R_{\mathrm{NT}}\right)
\,,  \end{gathered}
\end{equation}
where in the general case $\Delta z_{i}$ is the distance of the corresponding atom from the geometric center of the starting nanostructure along the $z$-axis (in our case $\Delta z_{i}=0$ for planar hBN monolayers). For ZZ (AC) nanotubes, the $x$-axis aligns with the AC (ZZ) direction and the $y$-axis with the ZZ (AC) direction, as shown in Fig. \ref{schema_nt}b (\ref{schema_nt}c). $R_{\mathrm{NT}}$ corresponds to the initial radius of the NT:
\begin{equation}\label{ntradiuseq}
     R_{\mathrm{NT}} = \frac{L_{\mathrm{sheet},y}}{2\pi} = \frac{n_{y}a_{y}}{2\pi}\,,
\end{equation}
which is equivalent to the common expression $\frac{a}{2\pi}\sqrt{n^{2}+nm+m^{2}}$ \cite{Dresselhaus2001} involving pairs $(n,m)$ of chiral indices, upon setting $(n,m)=(n_\mathrm{ZZ},0)$ for ZZ and $n=m=n_\mathrm{AC}$ for AC nanotubes. The initial length of the hBN nanotube is
\begin{equation}\label{ntlengthseq}
    L_{\mathrm{NT}} = L_{\mathrm{sheet},x} =  n_{x}a_{x}  \; \,.
\end{equation}
The examined NTs are periodic along the axial direction $x$; hence, they are considered infinitely long. Using this notation, $n_{x}$ corresponds to the number of unit cell replications along the axis of the NT and $n_{y}$ the replications along the circumference of the NT, see illustrations in Fig. \ref{schema_nt}.

It is noted that equations (\ref{ntradiuseq}) and (\ref{ntlengthseq}) are approximate for NTs with small radius due to residual stress. The nanotube structure was relaxed via a two-level optimization scheme:
\begin{enumerate}
  \item The inner level (energy minimization) entails minimizing the potential energy of a NT with a fixed (input) length and returning the stress along the axial direction (output). The minimization is performed sequentially using the Polak-Ribiere conjugate-gradient \cite{Polak1969} and the Hessian-free truncated Newton algorithms \cite{Thompson2022} until the norm of the global force vector is less than $10^{-9} \textrm{ ev}/$\AA.
  \item In the outer level (NT length optimization) the length of the NT is optimized until the stress of the minimized configuration (inner level) along the axial direction is zero. The optimization is performed with Brent's algorithm \cite{2020SciPy-NMeth, Brent1973}  until changes in NT length are smaller than $10^{-12}$ \AA. After the optimization, all stress tensor components become zero.
\end{enumerate}

\begin{figure}[h]
\centering
  \includegraphics[width=8.25cm]{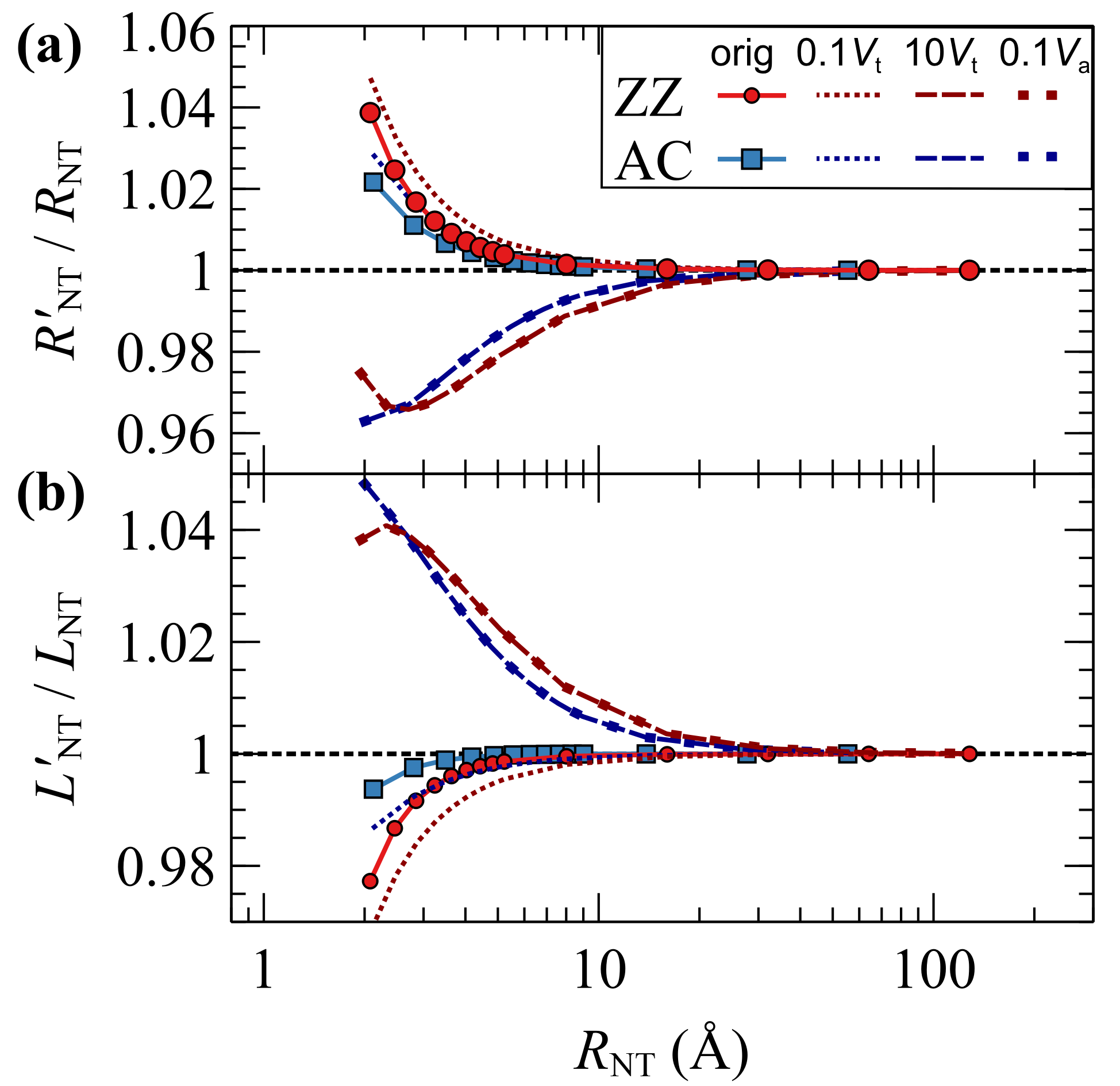}
  \caption{ Ratio of \textbf{(a)} the relaxed nanotube radius $R'_\textrm{NT}$ over the approximate value of Eq.~(\ref{ntradiuseq}) and \textbf{(b)} the relaxed nanotube length $L'_\textrm{NT}$ over the length of its planar counterpart given in Eq.~(\ref{ntlengthseq}), as a function of the initial NT radius. Red and blue colors correspond to ZZ and AC nanotubes, respectively. The circles and squares connected by solid lines represent evaluations using the original potential.
  Fine-dotted and dashed lines denote evaluations using a modified potential where the 
 strength of the torsional potential terms (Eqs. (\ref{torsionaleq1}) and (\ref{torsionaleq2})) is scaled by a factor of 0.1 and 10, respectively. Large dots (overlapping with the dashed lines) illustrate evaluations with a modified potential where the strength of the bond-angle-bending term (Eq. (\ref{bendeq1})) is scaled by 0.1.}
  \label{nt01}
\end{figure}

According to the results of these numerical simulations, shown in Fig.~\ref{nt01} by symbols connected with continuous lines, the radius $R'_\mathrm{NT}$ (length $L'_\mathrm{NT}$) of the relaxed hBN nanotubes is slightly larger (smaller) than its approximate value given in Eq.~(\ref{ntradiuseq}) (Eq.~(\ref{ntlengthseq})). The effect is slightly more pronounced in ZZ nanotubes. In any case for larger radius NTs, when $R_\mathrm{NT} > 20$ \AA~, Eqs.~(\ref{ntradiuseq}) and (\ref{ntlengthseq}) accurately represent the actual values.

The deviations of the actual, relaxed values of NT's radius and length (for smaller radius nanotubes) from the approximate expressions of Eqs.~(\ref{ntradiuseq}) and (\ref{ntlengthseq}) are determined by an interplay between the strength of the bond-angle-bending and torsional angle terms. To demonstrate this effect, we performed additional evaluations using modified force field parameters where the coefficients of the bond-angle-bending and torsional terms in Table \ref{modeltable} were scaled accordingly. \textit{Increasing} the strength of the torsional terms by an order of magnitude ($10V_{\mathrm{t}}$)\textemdash or equivalently, \textit{decreasing} the strength of the bond-angle-bending terms by ten times ($0.1V_{\mathrm{a}}$)\textemdash reverses this effect and results in NTs with smaller diameters and larger lengths (dashed and large-dotted lines in Figs. \ref{nt01}a and \ref{nt01}b, which give identical results). On the contrary, decreasing the strength of the torsional terms by an order of magnitude ($0.1V_{\mathrm{t}}$) slightly enhances the observed effect; see dotted lines in Figs. \ref{nt01}a and \ref{nt01}b. Note that the contribution of the bond-stretching term is zero; thus, scaling the bond-stretching potential by an arbitrary amount and subsequently relaxing the NT would leave the bond-angle-bending and torsional angles unaffected, resulting to the same potential energy. Therefore, the structural relaxation of the nanotubes depends strictly on the relative strength of the bond-angle-bending and torsional angle potentials. 

\begin{figure}
\centering
    \includegraphics[width=8.25cm]{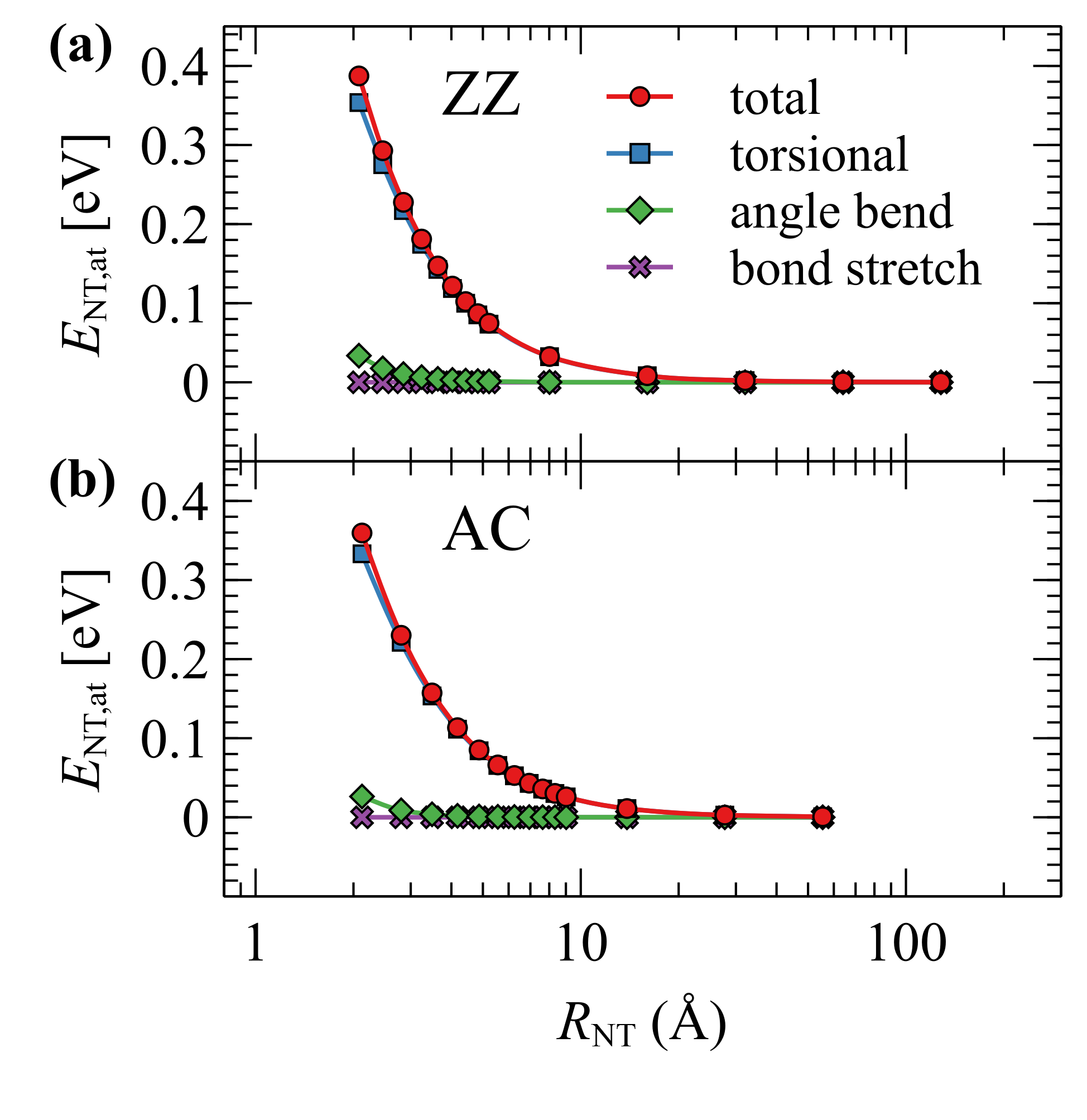}
  \caption{ Total energy per atom and the contribution of different potential energy terms in \textbf{(a)} ZZ and \textbf{(b)} AC nanotubes, as a function of the initial NT radius.}
  \label{nt02}
\end{figure}

Figures~\ref{nt02}a and \ref{nt02}b depict the variation of the potential energy per atom, $E_\mathrm{NT,at}$, for ZZ and AC nanotubes, respectively, as a function of the initial, non-relaxed NT's radius (circles connected by a red line). The separate contribution of each potential energy term of the force field is also shown. Bond-stretching contributions are shown by crosses, bond-angle-bending by diamonds, and torsional angles by squares.

The total deformation energy is dominated by the torsional terms which rise abruptly when the radius is decreasing, as expected. The contribution of bond-angle bending term is non-negligible only at very small radii. Bonds do not contribute to the potential energy of relaxed NTs.

\begin{figure}
\centering
    \includegraphics[width=8.25cm]{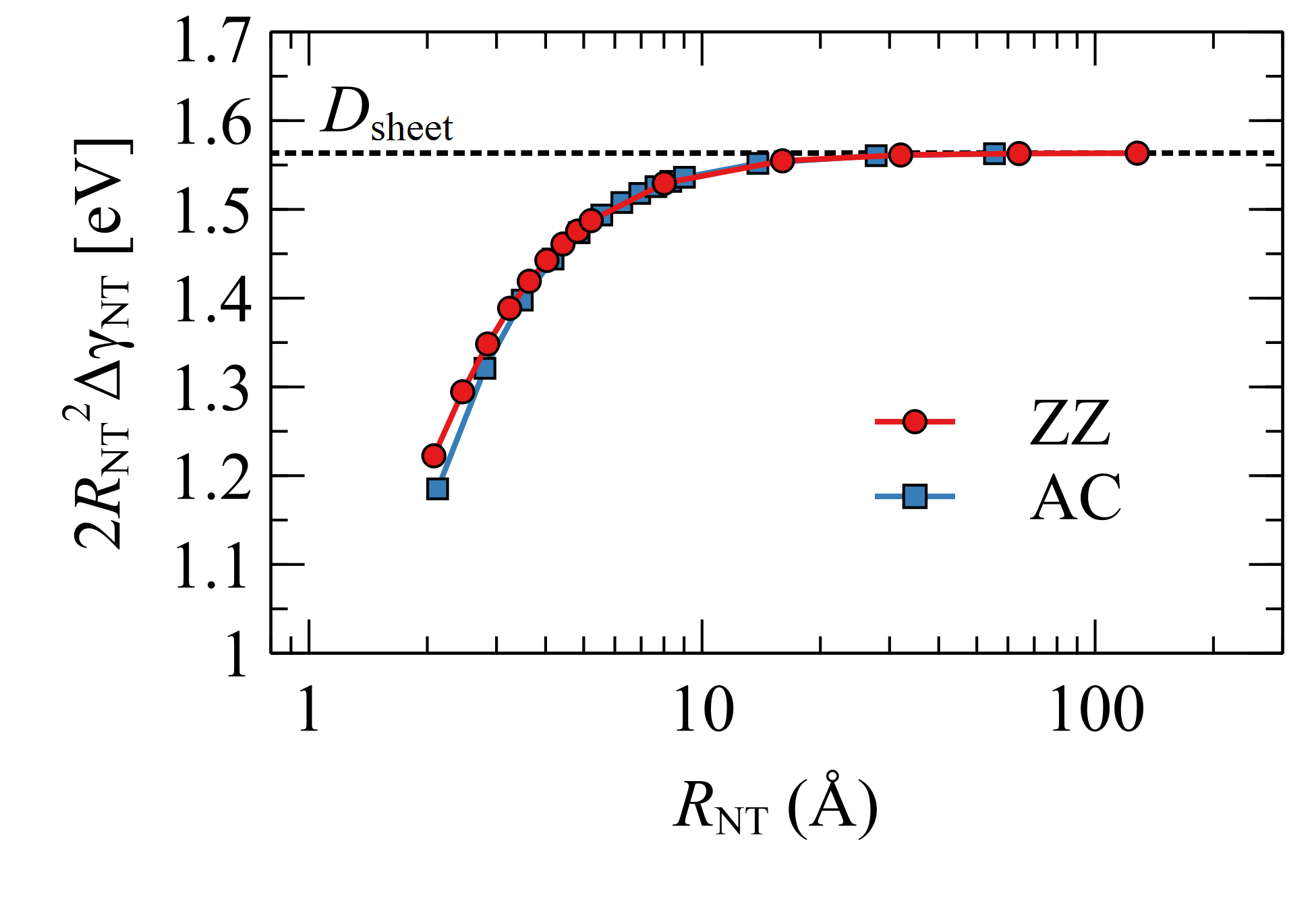}
  \caption{ Variation of the relative surface energy density $\Delta\gamma_\mathrm{NT}$ (Eq. (\ref{gammareleq})), scaled by the factor 2${R_\mathrm{NT}}^{2}$, with the NT radius for ZZ (circles) and AC (squares) nanotubes. Dashed line represents the analytically obtained bending rigidity ($D_\textrm{sheet} = 1.563$ eV) from Eq.~(\ref{bmeq2}).
  }
  \label{nt03}
\end{figure}

One can determine the bending rigidity of hBN monolayers as \cite{Wei2013}
\begin{equation}\label{bmeq0}
    D_\textrm{sheet}\equiv \frac{\partial^{2}\gamma_\mathrm{sheet}}{\partial \kappa^{2}}=
    \lim_{R_\mathrm{NT}\to\infty} 2{R_\mathrm{NT}}^{2}\Delta \gamma_\mathrm{NT}\,,
\end{equation}
with $\kappa$ being the curvature, and
\begin{equation}\label{gammareleq}
    \Delta\gamma_\mathrm{NT} = \gamma_\mathrm{NT} - \gamma_\mathrm{sheet} 
    = \frac{E_\textrm{NT,at}-E_\textrm{sheet,at}}{S_{0}}\,,
\end{equation}
the surface energy density of the NT ($\gamma_\mathrm{NT}=E_\textrm{NT,at}/S_{0}$) relative to that of the planar sheet ($\gamma_\mathrm{sheet}=E_\textrm{sheet,at}/S_{0}$), with $E_\textrm{NT,at}$ and $E_\textrm{sheet,at}$ being the per atom energy of the NT and sheet, respectively, and $S_{0} = (3\sqrt{3}/4){l_\textrm{BN}}^{2} = 2.738 \textup{~\AA}^{2}$ the surface area per atom.

The dependence of the quantity $2{R_\mathrm{NT}}^{2}\Delta\gamma_\mathrm{NT}$ on the NT radius is shown in Fig.~\ref{nt03} for ZZ and AC nanotubes. In both cases this quantity tends to the same value for large radii, providing the bending rigidity of hBN monolayer as indicated by Eq. (\ref{bmeq0}).

According to Davini et al. \cite{DAVINI201796} the bending rigidity of a single layer hBN can be determined analytically by the torsional potential energy terms, through the relation 
\begin{equation}\label{bmeq1}
  \begin{aligned}
    D_\textrm{sheet}&=\frac{2}{\sqrt{3}}\frac{d^{2}V_\mathrm{t}^\mathrm{cis}}{d\omega^{2}}\Bigr|_{\omega=0}
    +\frac{5}{\sqrt{3}}\frac{d^{2}V_\mathrm{t}^\mathrm{trans}}{d\omega^{2}}\Bigr|_{\omega=\pi} \\
    &-\frac{1}{2}\frac{dV_\mathrm{a}}{d\phi}\Bigr|_{\phi=2\pi/3}
    \,,
  \end{aligned}
\end{equation}
(compare with Eqs. (6) and (30) in Ref.~\onlinecite{DAVINI201796}). Evaluating the first derivative of Eq. (\ref{bendeq1}) at $\phi = 2\pi/3$ and the
second derivatives of Eqs.~(\ref{torsionaleq1}) and (\ref{torsionaleq2}) at $\omega = 0$ and $\omega = \pi$, respectively, Eq.~(\ref{bmeq1}) yields:
\begin{equation}\label{bmeq2}
    D_\textrm{sheet}=\frac{2}{\sqrt{3}} 4k_\mathrm{t}^\mathrm{cis}
                    +\frac{5}{\sqrt{3}}4k_\mathrm{t}^\mathrm{trans}
    \,,
\end{equation}
where the contribution of the last term in Eq. (\ref{bmeq1}) (self-stress) \cite{DAVINI201796} is zero because $\phi_{a}$ has been set to $2\pi/3$. The analytical value obtained from Eq. (\ref{bmeq2}) equals to $D_\mathrm{sheet} = 1.563 $~eV and it is depicted by the horizontal dashed line in Fig.~\ref{nt03}. It can be seen that the numerical estimation of bending rigidity from Eq.~(\ref{bmeq0}) in the limit of large NT radius coincides with the analytical estimate from Eq. (\ref{bmeq2}). For comparison, reported literature values of bending rigidity are shown in Table \ref{elastictable}.

Finally, the Gaussian stiffness of the hBN sheet can be determined analytically according to Ref. \cite{DAVINI201796}:
\begin{equation}\label{bgauss1}
    D_\textrm{G,sheet}=
    -\frac{8}{\sqrt{3}}\frac{d^{2}V_\mathrm{t}^\mathrm{cis}}{d\omega^{2}}\Bigr|_{\omega=0}
    -\frac{8}{\sqrt{3}}\frac{d^{2}V_\mathrm{t}^\mathrm{trans}}{d\omega^{2}}\Bigr|_{\omega=\pi}
    \,.
\end{equation}
By evaluating the second derivatives of Eqs.~(\ref{torsionaleq1}) and (\ref{torsionaleq2}) at $\omega = 0$ and $\omega = \pi$, Eq. (\ref{bgauss1}) becomes:
\begin{equation}\label{bgauss2}
    D_\textrm{G,sheet}=-\frac{8}{\sqrt{3}} (4k_\mathrm{t}^\mathrm{cis} + 4k_\mathrm{t}^\mathrm{trans})
    \,,
\end{equation}
which results to a Gaussian stiffness of -3.233 eV.
\subsection{\label{PeTorus}Elastic Energy of hBN Nanotori and the Bending Rigidity of hBN Nanotubes} 

\begin{figure}[h]
\centering
    \includegraphics[height=7.01cm,width=5.48cm]{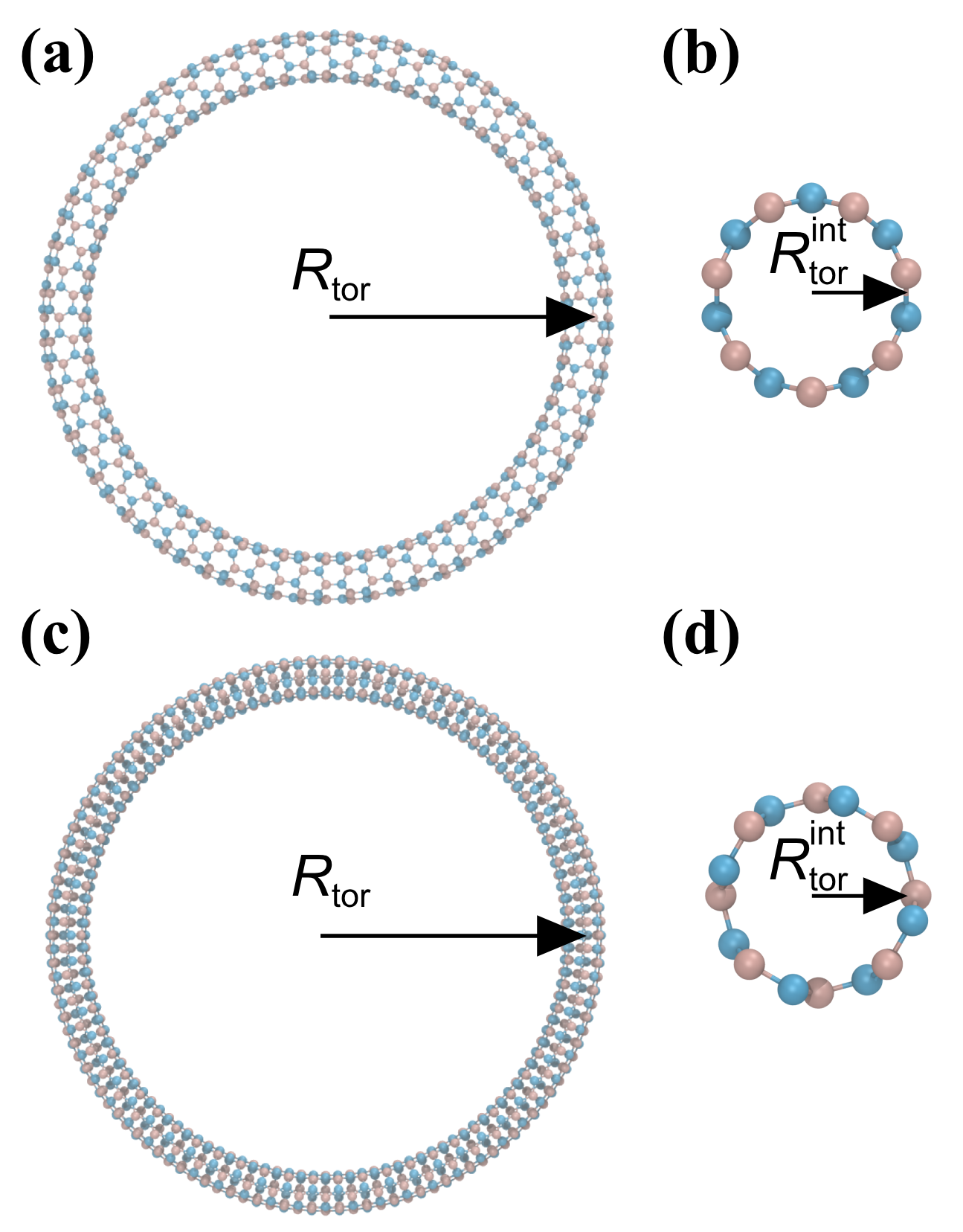}
  \caption{ Schematic representation of hBN nanotori. \textbf{(a)} A ZZ torus with external radius $R_\mathrm{tor} \approx 25$ \AA. The torus was formed by wrapping a  ZZ nanotube with length $L_\mathrm{NT} = 36a_{\mathrm{AC}} \approx 157 \text{~\AA}$ and radius $R_\mathrm{tor}^{\mathrm{int}} \approx R_\mathrm{NT} = 7a_{\mathrm{ZZ}}/(2\pi) \approx 2.8 \text{~\AA}$. \textbf{(b)} A cross-section of the wrapped ZZ nanotube in (a).
  \textbf{(c)} An AC torus with external radius $R_\mathrm{tor} \approx 25$ \AA, formed by wrapping an AC nanotube with length $L_\mathrm{NT} = 62a_{\mathrm{ZZ}} \approx 156 \text{~\AA}$ and radius $R_\mathrm{tor}^{\mathrm{int}} \approx R_\mathrm{NT} = 4{a_\mathrm{AC}}/(2\pi) \approx 2.8 \text{~\AA}$. \textbf{(d)} A cross-section of the wrapped AC nanotube in (c).}
  \label{torus_schema}
\end{figure}

We extend our energetic considerations in a variety of hBN nanotori with either ZZ or AC chiralities, in order to evaluate the bending rigidity of hBN nanotubes similarly to the procedure followed in the previous subsection. The initial configurations of the ZZ/AC tori were generated upon wrapping ZZ/AC nanotubes along their axial (periodic) direction by applying an analogous deformation scheme as in Eq.~(\ref{ntdeformeq}).
Figure~\ref{torus_schema} depicts a schematic illustration of the obtained nanostructures.

Each nanotorus is characterized by an internal $R_\mathrm{tor}^\mathrm{int}$ and an external $R_\mathrm{tor}$ radius. The internal radius can be approximately given by the radius of the nanotube that was wrapped to generate the torus:
\begin{equation}\label{Rtorus_int}
    R_\mathrm{tor}^\mathrm{int}=R_\mathrm{NT}\,.
\end{equation}
Similarly, the external torus radius can be approximated using the wrapped NT length, through the relation
\begin{equation}\label{Rtorus_ext}
R_\mathrm{tor}=\frac{L_\mathrm{NT}}{2\pi}\,.
\end{equation}

In the following, the total deformation energy of the tori configurations was minimized with the Polak-Ribiere conjugate-gradient algorithm \cite{Polak1969} followed by the Hessian-free truncated Newton optimizer \cite{Thompson2022} until the norm of the global force vector became less than $10^{-9} \textrm{ ev}/$\AA. As a result of this relaxation process, the obtained internal/external torus radii were slightly different than their values provided in Eqs.~(\ref{Rtorus_int}) and (\ref{Rtorus_ext}).

\begin{figure}[h]
\centering
    \includegraphics[width=8.25cm]{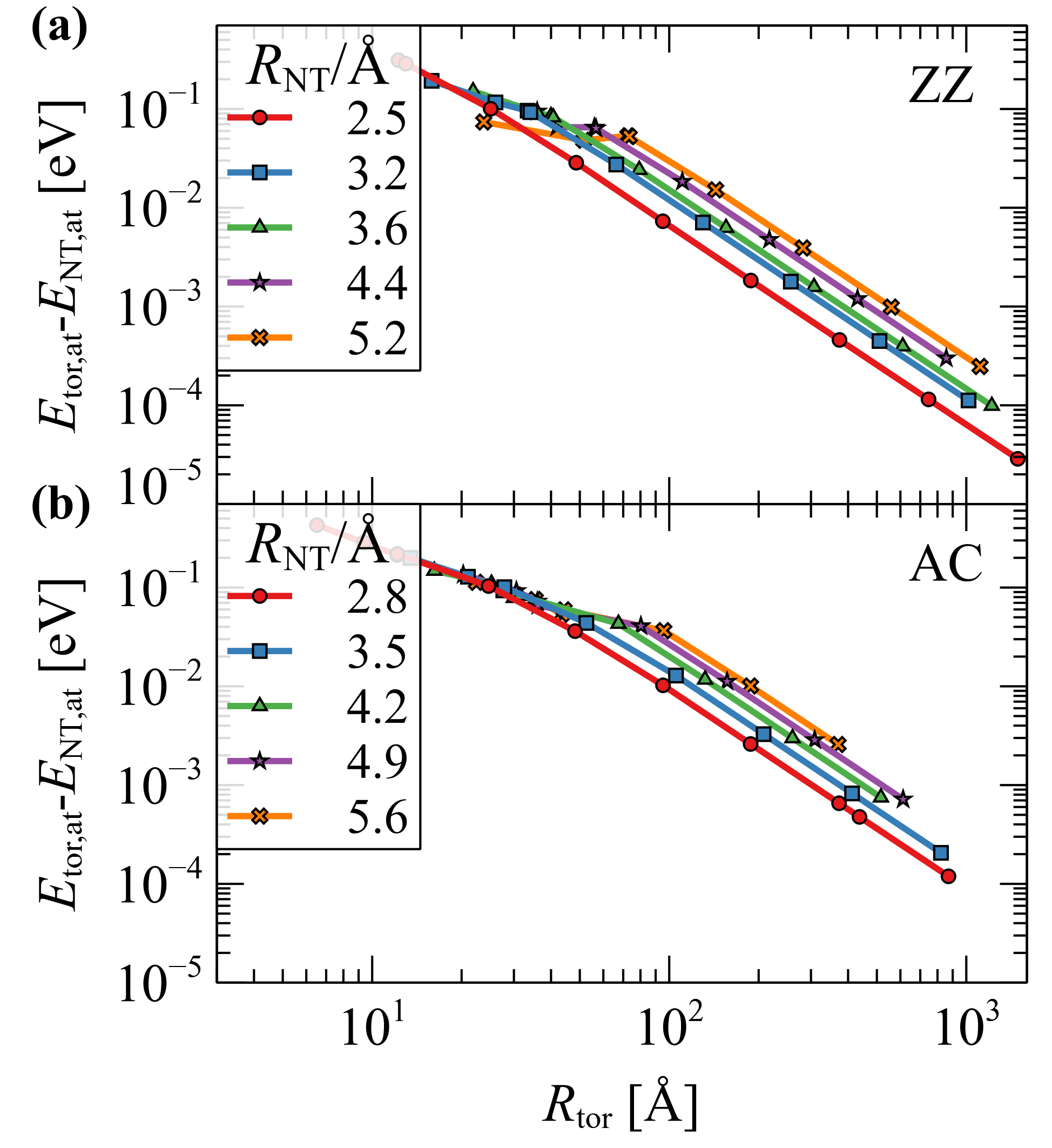}
  \caption{Total energy per atom of (a) ZZ and (b) AC hBN nanotori relative to that of the corresponding hBN nanotubes with the same internal radius and chirality, as a function of the torus external radius.
  Different colors represent nanotori of different internal radii, as indicated in the plots.}
  \label{torus_energy_01}
\end{figure}

Figures~\ref{torus_energy_01}a and \ref{torus_energy_01}b present the total energy per atom of relaxed ZZ and AC tori, with respect to the corresponding energy per atom of the nanotube with the same chirality and internal radius as the wrapped NT used to construct the torus. Note that, in the limit $R_\mathrm{tor} {\rightarrow} \infty$ the potential energy per atom of the torus equals the potential energy per atom of the NT; as a consequence, the energy difference in Figs.~\ref{torus_energy_01}a and \ref{torus_energy_01}b approaches zero.

There is a lower threshold for the external radius of the nanotorus, below which it becomes unstable and buckles,\cite{10.1063/1.4754538} resulting in the nonlinearities observed in Figures \ref{torus_energy_01}a and \ref{torus_energy_01}b.
For larger internal radii these instabilities arise at larger values of the external radius.

\begin{figure}[h]
\centering
    \includegraphics[width=8.25cm]{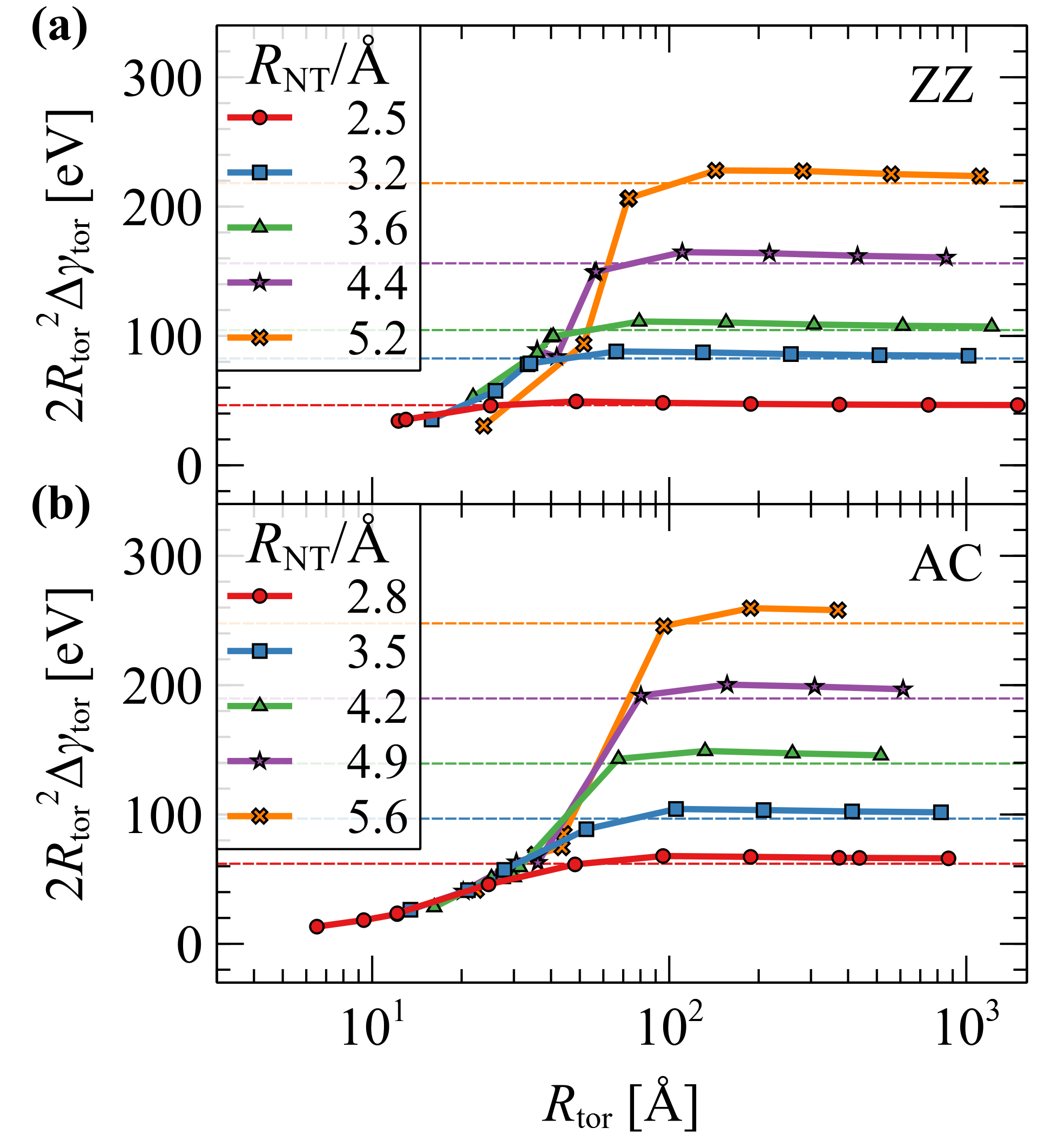}
  \caption{ Relative surface energy density $\Delta\gamma_\mathrm{tor}$, Eq. (\ref{gammatoruseq2}), multiplied by $2{R_\mathrm{tor}}^2$ for \textbf{(a)} ZZ and \textbf{(b)} AC hBN nanotori, as a function of the torus external radius. Different colors represent nanotori of different internal radii, as indicated in the plots. Dashed lines represent evaluations of the bending rigidity of the hBN nanotubes using Eq. (\ref{ring_energy3}).}
  \label{torus_energy_02}
\end{figure}

Similarly to the discussion of the previous subsection, one can determine the bending rigidity of hBN nanotubes with different radii ($R_\textrm{NT}$) and chiralities, by considering the corresponding tori in the limit of large external radius through the relation:
\begin{equation}\label{b_torus0}
    D_{\textrm{NT}}(R_\textrm{NT})\equiv \frac{\partial^{2}\gamma_\mathrm{NT}}{\partial \kappa^{2}}
    =\lim_{R_\mathrm{tor}\to\infty} 2{R_\mathrm{tor}}^{2}\Delta\gamma_\mathrm{tor}\,,
\end{equation}
with
\begin{equation}\label{gammatoruseq2}
    \Delta\gamma_\mathrm{tor} =
    \gamma_\mathrm{tor}-\gamma_\mathrm{NT}=
    \frac{E_\textrm{tor,at}-E_\textrm{NT,at}}{S_{0}}\,,
\end{equation}
being the surface energy density of the nanotorus with respect to that of a nanotube with the same chirality and the same diameter as the inner torus diameter.

It is noteworthy that, treating the NT as a one-dimensional object, an alternative definition for its bending rigidity \cite{Yakobson2001} can be expressed as:
\begin{equation}\label{gammatoruseq3}
    D'_\mathrm{NT} \equiv \frac{1}{L_\mathrm{NT}} \frac{\partial^{2}E_\mathrm{NT}}{\partial\kappa^{2}} = 2\pi R_\mathrm{NT} D_{\mathrm{NT}}\,,
\end{equation}
with energy$\times$length units.

Numerical calculations of the quantity $2{R_\mathrm{tor}}^{2}\Delta\gamma_\mathrm{tor}$ as a function of $R_\mathrm{tor}$ are shown in Figs.~\ref{torus_energy_02}a and \ref{torus_energy_02}b for ZZ and AC nanotori, respectively, for different values of the internal radius. The bending rigidity of the corresponding nanotubes can be obtained in the limit $R_\mathrm{tor}\to\infty$, according to Eq. (\ref{b_torus0}). As expected, the bending rigidity of nanotubes is an increasing function of their internal radius.

According to Ref.~\cite{10.1063/1.4754538}, the elastic strain energy for bending a hollow cylinder (thickness $l_\mathrm{t}$, radius $R_\mathrm{NT}$) into a ring of radius $R_\mathrm{tor}$ is
\begin{equation}\label{ring_energy1}
    E_\textrm{tor}-E_\textrm{NT}
        =\frac{{\pi}^{2}YR_\mathrm{NT}}{4R_\mathrm{tor}}\left( (2R_\mathrm{NT})^{2} + {l_\mathrm{t}}^{2} \right)\,,
\end{equation}
where we have adapted the notation of the present paper. By dividing Eq. (\ref{ring_energy1}) with the area of a torus, $S_\textrm{tor}~=~4\pi^{2}R_\mathrm{NT}R_\mathrm{tor}$, we can quantify the variation of the relative surface energy density as follows:
\begin{equation}\label{ring_energy2}
    \Delta\gamma_\textrm{tor}
      =Y
      \left[
        {\frac{1}{4}\left({\frac{R_\mathrm{NT}}{R_\mathrm{tor}}}\right)^2}  +
        \frac{1}{16}{\left({\frac{l_\mathrm{0}}{R_\mathrm{tor}}}\right)^2}
      \right] \,.
\end{equation}
The corresponding thickness $l_\mathrm{t}$ of the monolayer hBN sheet can be approximatelly ignored, therefore, in this limit Eq. (\ref{ring_energy2}) reads
\begin{equation}
\Delta\gamma_\textrm{tor}=\frac{Y}{4}\left({\frac{R_\mathrm{NT}}{R_\mathrm{tor}}}\right)^2 \,.
\end{equation}
Substituting last equation in Eq. (\ref{b_torus0}) yields
\begin{equation}\label{ring_energy3}
D_\mathrm{NT}(R_\mathrm{NT}) = \frac{Y}{2}R_\mathrm{NT}^2
\end{equation}
Eq. (\ref{ring_energy3}) provides an analytical expression for the bending rigidity of the hBN nanotubes with relatively large diameters, where $D_\textrm{NT}$ scales proportionally to the square of the nanotube radius.

Evaluations of Eq. (\ref{ring_energy3}) are illustrated with dashed lines in Fig. \ref{torus_energy_02}, for each value of the internal torus radii considered therein. In the limit of large $R_\mathrm{rot}$, the numerical evaluations of the bending rigidity ($2{R_\mathrm{tor}}^{2}\Delta\gamma_\mathrm{tor}$) are in excellent agreement with the analytical estimation from Eq.~(\ref{ring_energy3}). Surprisingly, the variation of $2{R_\mathrm{tor}}^{2}\Delta\gamma_\mathrm{tor}$ across the stable $R_\mathrm{tor}$ regime does not monotonically approach its limiting value corresponding to the NT bending rigidity. Instead, it exhibits a maximum as $R_\mathrm{tor}$ increases before asymptotically approaching the limiting constant from above.

\section*{\label{Conclusions}Conclusions}

We develop an atomistic force field for hexagonal boron nitride nanostructures including 2D monolayers, single-layer nanotubes, and nanotori. The force field is parameterized via a systematic bottom-up procedure calibrated through first principle’s calculations (DFT) of properly deformed hBN monolayers, ensuring accurate representation of both in-plane and out-of-plane deformations. The parameterization included data from significantly deformed sheets accounting for nonlinear effects, but it did not extend to extreme deformations. Additional density functional theory calculations, performed using Gaussian16\cite{Gaussian16} on hydrogen-passivated BN flakes (results not shown), revealed that the electronic character of hBN remains largely unaffected by out-of-plane bending. This stability is attributed to the robust electronic structure, as evidenced by the large HOMO-LUMO gap.

The components of the stiffness matrix and the corresponding elastic constants (i.e., Young's modulus, Poisson's ratio, and bulk modulus) were calculated via a finite difference scheme using molecular mechanics calculations and validated against analytical formulas. The predicted mechanical properties are in agreement with experimental and theoretical values reported in the literature. For the potential energy terms considered in our force field model, the Poisson’s ratio of the nanosheets depends strictly on the relative strength of the bond-stretching and angle-bending interactions.

Nanotubes with both AC and ZZ chiralities were investigated over a broad range of NT radii. Small-radius nanotubes exhibited noticeable structural relaxation, which is determined by the relative strength of torsional and bond-angle-bending terms of the potential. 
The bending rigidity of hBN sheets was estimated through the limiting value of the total deformation energy of NTs at increasingly large radii; the obtained numerical estimate is in excellent agreement with an analytical formula based on the coefficients of the torsional terms of the force field.

The elastic energy of carbon nanotori has been successfully described using a continuum model for bent hollow beams \cite{10.1063/1.4754538}. Building on this framework, by applying the fundamental definition of bending rigidity to the continuum model we derived a model for predicting the bending rigidity of nanotubes. To assess the accuracy of our model, we compared its predictions with numerical estimations of bending rigidity of hBN nanotubes with a given radius, obtained through the deformation energy of the corresponding nanotori with the same internal radius in the limit of very large external radii. We found that the nanotube bending rigidity increases quadratically with its radius and proportionally with Young's modulus, in agreement with our model. Interestingly, while the relative surface energy density follows a ${R_\mathrm{tor}}^{-2}$ scaling at large radii, its behavior becomes non-monotonic for smaller radii above the critical buckling point. Within this regime, $2{R_\mathrm{tor}}^{2}\Delta\gamma_\mathrm{tor}$ reaches a maximum with increasing ${R_\mathrm{tor}}$ before asymptotically approaching its limiting value (associated with nanotube bending rigidity) from above.

The proposed force field belongs to the class of classical, non-reactive force fields, sharing the advantages of simplicity, ease of use, and efficiency, similar to relevant models in the literature \cite{nano11113113, Boldrin_2011, GovindRajan2018, Mayo1990}. A key distinction of our implementation, however, lies in the incorporation of nonlinear terms in the in-plane bond-stretching and bond-angle-bending potentials. These terms account for anharmonic effects, which become significant at large deformations and/or elevated temperatures.

As demonstrated in Ref \cite{10.1063/1.4798384} through a direct comparison between nonlinear and linearized potentials for graphene, the inclusion of the nonlinear terms considerably improves the accuracy of the predicted elastic response under large uniaxial extensions; see Fig. 7, therein. The linearized potentials become very inaccurate beyond strain values of 0.05 where the anharmonic effects become important, Poisson's ratio is particularly sensitive to anharmonic effects as its qualitative behavior changes significantly in their absence \cite{10.1063/1.4798384}. In addition, linearized force fields fail to capture phonon softening—the redshift of the phonon density of states—that occurs with increasing strain due to weakened interatomic interactions.

Tersoff-based potentials \cite{Thomas_2016, PhysRevB.87.184106, YI2018408, MORTAZAVI20121846, Han_2014} and their extensions (e.g., the ExTeP force field \cite{PhysRevB.96.184108}), also employ nonlinear functions to model interatomic interactions, similar to our method. This enables them to qualitatively capture the elastic behavior of h-BN under large mechanical strains as well. However, the reactive and bond-order character of these potentials requires continuous monitoring of the local coordination number and potential reactions. These processes involve constructing neighboring lists during the simulations, which increases the communication overhead and makes them computationally expensive. Indeed, according to benchmarks (results not shown in the Arxiv version), the proposed force field is roughly five times faster than a Tersoff-based potential \cite{PhysRevB.96.184108}.

An additional utility of the class of nonreactive force fields lies in the ease with which their parameters can be related to the resulting model behavior. Specifically, simple closed-form expressions enable the exact calculation of both in-plane and out-of-plane elastic constants at 0 K through Eqs. (\ref{lambdaanalytseq})-(\ref{kappatildeeq}) and (\ref{bmeq1})-(\ref{bgauss1}), respectively. These analytical expressions are generalizable to any 2D membrane model that shares the same crystalline structure and consists solely of bonds, bond angles, and dihedral angles. This feature is particularly valuable for top-down parameterization of such 2D membranes, especially in mesoscopic coarse-grained modeling. For example, if the in-plane and out-of-plane elastic constants are known, the corresponding force field parameters can be determined analytically, ensuring the reproduction of the target elastic properties at 0 K.

To summarize, the developed force field is simple, numerically efficient, yet effective, capable of accurately capturing the mechanical properties of hBN nanostructures in both equilibrium and nonlinear regimes. Its versatility could be useful for the development of hBN-based materials in applications such as thermal management, optoelectronics, and energy storage, among others.

\section*{\label{data_availability_statement}Data Availability Statement}
The data that support the findings of this article are openly available \cite{article_data}.



\bibliography{apssamp}

\end{document}